\documentclass[a4paper,11pt]{article}

\pdfoutput=1 

\usepackage{jheppub,bm,slashed} 
\usepackage[legacycolonsymbols]{mathtools}
\usepackage[usenames,dvipsnames,svgnames,table,x11names]{xcolor}
\usepackage{extarrows}
\usepackage{easybmat}

\usepackage{nicematrix}

\usepackage[T1]{fontenc} 
\usepackage[utf8]{inputenc}

\renewcommand*{\backref}[1]{}
\renewcommand*{\backrefalt}[4]{%
  \ifcase #1 %
    \relax 
  \or
    {\scriptsize (page~#2).}%
  \else
    {\scriptsize (pages~#2).}%
  \fi%
}
\newcommand{\minus}{\text{-}}

\definecolor{light_blue}{rgb}{0.15, 0.35, 0.95}
\definecolor{kit_green}{rgb}{0, 
0.58823 
, 0.50980 
}

\usepackage{tikz}
\usepackage{pgfplots}
\pgfplotsset{compat=1.14}
\usepackage{tikz-3dplot}
\usetikzlibrary{intersections, calc,positioning,decorations.pathreplacing,decorations.pathmorphing,arrows,arrows.meta,patterns,pgfplots.fillbetween,math,matrix}

\usepackage{newtxtext}
\usepackage{bm}
\usepackage{amsmath}
\usepackage{amsfonts}
\usepackage{amssymb}
\usepackage{latexsym}
\usepackage{graphicx}
\usepackage{float}
\usepackage{siunitx}
\usepackage{booktabs}
\usepackage{braket}
\usepackage{physics}
\usepackage{cancel}
\usepackage{ascmac}
\usepackage{comment}

\newcommand{\C}{\mathbb{C}}

\newcommand{\R}{\mathbb{R}}

\newcommand{\Z}{\mathbb{Z}}

\newcommand{\mbf}{\mathbf}

\newcommand{\Spin}{\mathrm{Spin}}
\newcommand{\SO}{\mathrm{SO}}

\newcommand{\SU}{\mathrm{SU}}
\newcommand{\U}{\mathrm{U}}

\newcommand{\Sp}{\mathrm{Sp}}

\usepackage{cancel}

\usepackage{nicematrix}

\usepackage[whole]{bxcjkjatype} 

\usepackage[T1]{fontenc} 
\usepackage[utf8]{inputenc}

\renewcommand*{\backref}[1]{}

\definecolor{light_blue}{rgb}{0.15, 0.35, 0.95}
\definecolor{kit_green}{rgb}{0, 
0.58823 
, 0.50980 
}

\usepackage{tikz}
\usepackage{pgfplots}
\pgfplotsset{compat=1.14}
\usepackage{tikz-3dplot}
\usetikzlibrary{intersections, calc,positioning,decorations.pathreplacing,decorations.pathmorphing,arrows,arrows.meta,patterns,pgfplots.fillbetween,math,matrix}

\usepackage{multirow}
\usepackage{booktabs} 
\usepackage{siunitx} 
\usepackage{tabularx} 

\preprint{
\begin{minipage}{5cm}
\flushright
KEK-TH-2723\\KYUSHU-HET-323
\end{minipage}}

\title{More on 8d non-supersymmetric branes and heterotic strings}

\author[1,2]{Yuta Hamada,}
\author[2]{Arata Ishige}
\author[3]{and Yuichi Koga}
\affiliation[1]{Theory Center, IPNS, High Energy Accelerator Research Organization (KEK), 1-1 Oho, Tsukuba, Ibaraki 305-0801, Japan}
\affiliation[2]{Graduate Institute for Advanced Studies, SOKENDAI, 1-1 Oho, Tsukuba, Ibaraki 305-0801, Japan}
\affiliation[3]{Institute for Advanced Study, Kyushu University, 744 Motooka, Nishi-ku, Fukuoka 819-0395, Japan}

\emailAdd{yhamada@post.kek.jp}
\emailAdd{arata@post.kek.jp}
\emailAdd{koga.yuichi@phys.kyushu-u.ac.jp}

\abstract{
We determine the maximal gauge groups arising in $E_8$ heterotic string theory on $T^2$.
Our analysis proceeds along two approaches.
First, we start the moduli space of the supersymmetric heterotic string theory on $T^2$ focusing on points of maximal gauge enhancement. 
At these special points, the charge lattice can exhibit a $\mathbb{Z}_2$ outer automorphism corresponding to the bulk gauge symmetry.
By orbifolding the worldsheet theory by it with the fermion parity, we obtain the maximal gauge group of the $E_8$ theory.
Second, we directly study the toroidal compactification of 10d $E_8$ heterotic string.
Both approaches agree, yielding a classification of $22$ maximal gauge groups.
For each case, we present the corresponding massless spectrum. 
In light of the no global symmetry/cobordism conjecture in quantum gravity, our result also offer a classification of non-supersymmetric branes in 8d supersymmetric heterotic string theories.
}

\begin{document}

\maketitle

\section{Introduction}

In the context of heterotic strings~\cite{Gross:1984dd}, it is known that there are two supersymmetric theories ($E_8\times E_8 $ and $SO(32)$) and seven non-supersymmetric theories ($SO(16)\times SO(16),SO(16)\times E_8,SO(32), (E_7\times SU(2))^2, U(1)\times SU(16)$, $SO(8)\times SO(24), E_8$) in ten dimensions \cite{Harvey:1996ur,Alvarez-Gaume:1986ghj,Kawai:1986vd}. 
Recently, it is shown that there are no unknown 10d heterotic string theories~\cite{BoyleSmith:2023xkd,Rayhaun:2023pgc,Hohn:2023auw}.
Among non-supersymmetric theories, only the $SO(16)\times SO(16)$ theory is non-tachyonic, and only the $E_8$ theory has the reduced-rank gauge symmetry.

The compactification has been well-investigated in light of the Swampland conjectures~\cite{Vafa:2005ui} as well as the string lampposts/string universality~\cite{Adams:2010zy,Kim:2019vuc,Kim:2019ths,Cvetic:2020kuw,Montero:2020icj,Hamada:2021bbz,Bedroya:2021fbu,Cvetic:2022uuu}.
For instance, the toroidal compactifications of the supersymmetric heterotic string and CHL string theories are studied in \cite{Font:2020rsk,Font:2021uyw,Fraiman:2021soq,Fraiman:2021hma}.

\begin{figure}[htb]

\begin{center}
\begin{tikzpicture}[
  node distance=2.5cm and 3cm,
  every node/.style={align=center},
  arrow/.style={-{Stealth}, thick},
  dashedarrow/.style={-{Stealth}, thick, dashed},
  labelstyle/.style={font=\footnotesize}
]

\node (SUSY) at (-3,4.5) {SUSY};
\node (10dleft) at (-3,4) {$E_8 \times E_8$};
\node (non-SUSY) at (3,4.5) {\cancel{SUSY}};
\node (10dright) at (3,4) {$E_8$};

\node (9dleft) at (-3,2) {$2E_{9-n} \times A_{2n-1}$};
\node (9dright) at (3,2) {$E_{9-n} \times C_n$};

\node (8dleft) at (-3,0) {Various\\Gauge Groups};
\node (8dright) at (3,0) {Various\\Gauge Groups};

\node (10d) at (-5,4) {10d};
\node (9d) at (-5,2) {9d};
\node (8d) at (-5,0) {8d};
\draw[arrow] (10dleft) -- (10dright) node[midway, above, labelstyle] {\cite{Kawai:1986vd}};
\draw[arrow] (10dleft) -- (9dleft) node[midway, right, labelstyle] {\cite{Cachazo:2000ey,Font:2020rsk}};
\draw[arrow] (9dleft) -- (9dright) node[midway, above, labelstyle] {\cite{Hamada:2025JHEP}};
\draw[arrow] (10dleft) to[bend right=25] node[pos=0.8, right, labelstyle]{\cite{Font:2020rsk}} (8dleft) ; 
\draw[arrow,draw=red] (8dleft) -- (8dright) node[midway, below, labelstyle,text=red] {New};

\draw[arrow,draw=red] (10dright) -- (9dright) node[midway, right, labelstyle,text=red] {New};
\draw[arrow,draw=red] (10dright) to[bend left=45] (8dright);

\end{tikzpicture}
 \caption{A schematic picture of the relation between the 10d, 9d, and 8d heterotic theories.}
    \label{fig:relation}
\end{center}

\end{figure}
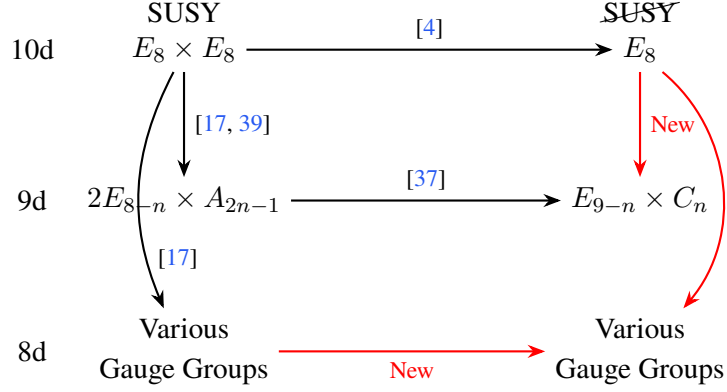

Regarding the non-supersymmetric strings, the $S^1$ compactification of non-supersymmetric heterotic string theory is studied in \cite{Fraiman:2023cpa}.\footnote{
    See also {\textit e.g.}, for \cite{Blaszczyk:2014qoa,Hamada:2015ria,Ashfaque:2015vta,Blaszczyk:2015zta,Itoyama:2019yst,Itoyama:2020ifw,Itoyama:2021fwc,Itoyama:2021itj,Koga:2022qch,Nakajima:2023zsh,Saxena:2024eil,Baykara:2024tjr,Detraux:2024esd,Abel:2024vov,Escalante-Notario:2025hvn,Funakoshi:2025lxs} for recent studies on non-supersymmetric heterotic strings. 
}
Moreover, two of the authors constructed nine-dimensional non-supersymmetric heterotic strings with reduced-rank gauge symmetry \cite{Hamada:2025JHEP}. 
There, we obtain all of the possible nine-dimensional theories but only a few eight-dimensional examples.
In this paper, building on the previous study, we construct all possible maximal gauge enhancements arising in $E_8$ heterotic string theory on $T^2$.
This is achieved by two approaches (see figure~\ref{fig:relation}).
The first approach is similar to the one in \cite{Hamada:2025JHEP}, where we start from the moduli space of the supersymmetric heterotic string theory on $T^2$ and focus on points of maximal gauge enhancement~\cite{Font:2021uyw}.
Then, we find that the charge lattice sometimes exhibits a $\mathbb{Z}_2$ outer automorphism corresponding to the bulk gauge symmetry.
When this happens, we can obtain the maximal gauge group of the $E_8$ theory by orbifolding the worldsheet theory by this outer automorphism and the spacetime fermion parity.
We also investigate toroidal compactifications of the $E_8$ heterotic string theory directly and determine the moduli that lead to the nine- and eight-dimensional theories with reduced-rank gauge symmetries. 
Furthermore, we propose the generalized Dynkin diagrams for the $E_8$ string theory that can be used to identify the maximally enhanced theories. 
Using the no global symmetry conjecture or the cobordism conjecture~\cite{McNamara:2019rup} of the quantum gravity, our classification can be interpreted as a classification of the non-supersymmetric codimension-two branes in supersymmetric heterotic string theory in 8d.

The paper is organized as follows.
In section \ref{sec:orbifold}, we review the asymmetric orbifolding of the heterotic string theory by the folding of Dynkin diagrams.
In section \ref{sec:9d_review}, we review our previous results in 9d~\cite{Hamada:2025JHEP}.
In section \ref{sec:8d_folding}, we perform the analysis of the eight-dimensional theories obtained by the asymmetric orbifolding of the supersymmetric theories.
In section \ref{sec:E8}, we discuss the direct construction of the eight-dimensional theories from the ten-dimensional $E_8$ heterotic string theory on $T^2$.

\section{Asymmetric Orbifolding of Heterotic string theory}\label{sec:orbifold}
In this section, we review the asymmetric orbifolding of the heterotic string theory and the folding of ADE Dynkin diagrams.
\subsection{Construction}
The starting point is the torus partition function of $(10-d)-$dimensional supersymmetric heterotic string theory given by
\begin{equation}
\begin{aligned}
    Z^{(10-d)}_{\text{SUSY}}(\tau,\bar{\tau})
         =&Z_B^{(8-d)} \qty(\bar{V}_8-\bar{S}_8) \frac{1}{\eta^{16+d}\bar{\eta}^d}\sum_{p\in\Gamma_{16+d,d}}q^{\frac{1}{2}p_L^2}\bar{q}^{\frac{1}{2}p_R^2},
\label{eq:SUSY_partition_function}\end{aligned}
\end{equation}
where the $(16+d,d)$ lattice $\Gamma_{16+d,d}$ is even and self-dual, so-called Narain lattice, $\tau\in \C$ with $\Im \tau>0$ is a complex modulus of a torus, and $q=e^{ 2\pi i \tau }$.  We used  
\begin{equation}
    Z_B^{(8-d)}:=\frac{1}{(\Im \tau)^{\frac{8-d}{2}}}\frac{1}{(\eta\bar{\eta})^{8-d}},\\
\label{eq:ZB}
\end{equation}
where $\eta$ is the Dedekind eta function, and $V_8, S_8$ are the $D_4$ character vector and spinor conjugacy classes defined in Appendix \ref{sec:theta}.

If the lattice $\Gamma_{16+d,d}$ admits a $\Z_2$ symmetry $g$, then a $\Z_2$ orbifolding can be discussed using its symmetry. One can construct the $(8+d,d)$ lattice $I_{8+d,d}$ by the action of $g$ on $\Gamma_{16+d,d}$ as follows:
\begin{equation}
    I_{8+d,d}\coloneqq\{ p\in\Gamma_{16+d,d} | g(p)=p\},
\end{equation}
which is called invariant lattice. A new theory obtained by this orbifolding is based on $I_{8+d,d}$, and this process is called asymmetric orbifolding~\cite{Narain:1986qm} because a symmetry $g$ acts only left part of the lattice.

To obtain the non-supersymmetric string theories, we consider the asymmetric orbifolding by $g(-1)^F$, where $F$ is the spacetime fermion number.
The partition function of the untwisted sector is given by
\begin{equation}
    Z^{\mathrm{(untwisted)}}=\frac{1}{2}Z_B^{(8-d)}\qty\big((\bar{V}_8-\bar{S}_8)Z(1,1)+(\bar{V}_8+\bar{S}_8)Z(1,g)),
\end{equation}
where
\begin{equation}
    \begin{aligned}
         Z(1,1)\coloneqq&
         \frac{1}{\eta^{16+d}\bar{\eta}^{d}}\sum_{p\in \Gamma_{16+d,d}}q^{\frac{1}{2}p_R^2}\bar{q}^{\frac{1}{2}p_R^2},\\
        Z(1,g)\coloneqq&\frac{1}{\eta^{16+d}\bar{\eta}^d}\qty(\frac{\eta^3}{\theta_2})^4\sum_{p\in I_{8+d,d}}q^{\frac{1}{2}p_R^2}\bar{q}^{\frac{1}{2}p_R^2}.
    \end{aligned}
\end{equation}
The modular invariance of the partition function requires the twisted sectors should be added. Using the Poisson summation formula and the modular transformation property of theta functions (see Appendix \ref{sec:theta}), we obtain
\begin{equation}
    Z^{\mathrm{(twisted)}}=\frac{1}{2}Z_B^{(8-d)}\qty((\bar{O}_8-\bar{C}_8)Z(g,1)-(\bar{O}_8+\bar{C}_8)Z(g,g)),
    \label{eq:twisted_partition}
\end{equation}
where
\begin{equation}
    \begin{aligned}
        Z(g,1)=&\frac{1}{\eta^{16+d}\bar{\eta}^d} \qty(\frac{\eta^3}{\theta_4})^4 \sum_{p\in I_{8+d,d}^\ast}q^{\frac{1}{2}p_L^2}\bar{q}^{\frac{1}{2}p_R^2},\\
        Z(g,g)=&-\frac{1}{\eta^{16+d}\bar{\eta}^d} \qty(\frac{\eta^3}{\theta_3})^4 \sum_{p\in I_{8+d,d}^\ast}q^{\frac{1}{2}p_L^2}\bar{q}^{\frac{1}{2}p_R^2}e^{\pi i p^2}.\\
    \end{aligned}
\end{equation}
Note that $p^2=p_L^2-p_R^2$. Here $O_8, C_8$ are the $D_4$ character scalar and co-spinor conjugacy classes defined in Appendix \ref{sec:theta} and $I_{8+d,d}^\ast$ is the dual lattice of $I_{8+d,d}$. For the modular invariance, $Z(g,1)$ should be invariant under $\tau\to \tau+2$, so every element $(p_L,p_R)\in I^\ast_{8+d,d}$ satisfies $p^2 \in\Z$. Then it holds that $I^\ast_{8+d,d}\subset I_{8+d,d}/2$, since $I_{8+d,d}\subset \Gamma_{18,2}$ is an even lattice.

Adding all of these together, the partition function of the non-supersymmetric theory constructed by this orbifolding is given as follows:
\begin{equation}
\begin{aligned}
    Z^{(10-d)}_{\cancel{\text{SUSY}}}=&\frac{1}{2}Z^{(8-d)}_B\left((\bar{V}_8-\bar{S}_8)Z(1,1)+(\bar{V}_8+\bar{S}_8)Z(1,g)\right.\\
    +& \left.(\bar{O}_8-\bar{C}_8)Z(g,1)-(\bar{O}_8+\bar{C}_8)Z(g,g)\right).
\end{aligned}
\end{equation}

\subsection{Folding of Dynkin Diagram}
As an action $g$, we consider the outer automorphism of the Lie algebra, which corresponds to the folding of Dynkin diagrams.
Here, we introduce the foldings relevant to our analysis.
\subsubsection{\texorpdfstring{$A_{2n-1}\to C_n$}{A2n-1 -> Cn}}
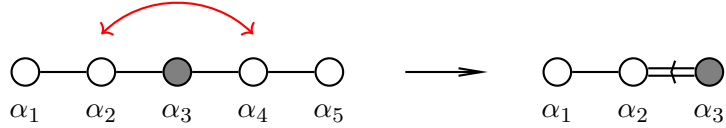
\begin{figure}[htb]
\centering
\begin{tikzpicture}[scale=1.0]

\node[thick, circle, draw, fill=white] (-1) at (-2,0) {};
\node[thick, circle, draw, fill=white] (0) at (-1,0) {};
\node[thick, circle, draw, fill=gray] (1) at (0,0) {};
\node[thick, circle, draw, fill=white] (2) at (1,0) {};
\node[thick, circle, draw, fill=white] (3) at (2,0) {};

\draw[thick] (-1) -- (0);
\draw[thick] (0) -- (1);
\draw[thick] (1) -- (2);
\draw[thick] (2) -- (3);

\node[below=0.3cm] at (-1) {$\alpha_1$};
\node[below=0.3cm] at (0) {$\alpha_2$};
\node[below=0.3cm] at (1) {$\alpha_3$};
\node[below=0.3cm] at (2) {$\alpha_4$};
\node[below=0.3cm] at (3) {$\alpha_5$};

\node[thick, circle, draw, fill=white] (1) at (5,0) {};
\node[thick, circle, draw, fill=white] (2) at (6,0) {};
\node[thick, circle, draw, fill=gray] (3) at (7,0) {};

\draw[thick] (1) -- (2);
\draw[thick] (3,0.0)--(4,0);
\draw[thick] (3.7,0.05)--(4,0)--(3.7,-0.05);

\draw[double distance=2pt, thick] (2) -- (3);
\draw[thick] (6.55,0.15) -- (6.5,0) -- (6.55,-0.15);
   \draw[thick, red, <->] (-1,0.5) to [out=45,in=135] (1,0.5);
\node[below=0.3cm] at (1) {$\alpha_1$};
\node[below=0.3cm] at (2) {$\alpha_2$};
\node[below=0.3cm] at (3) {$\alpha_3$};

\end{tikzpicture}
\caption{Folding $A_{2n-1}$ into $C_n$}
\label{Fig_AtoC}
\end{figure}

As an example, we describe the folding $A_{2n-1}$ into $C_n$. Let us consider the root lattice of $A_{2n-1}$:
\begin{equation}
    \Lambda_R(A_{2n-1})=\bigoplus_{i=1}^{2n-1}\Z\alpha_i^{(A_{2n-1})}.
\end{equation}
The specific form of $\alpha_i^{(A_{2n-1})}$ is summarized in Appendix \ref{sec:lie algebra}. This lattice admits a $\Z_2$ symmetry $g$ which flips $\alpha_i^{(A_{2n-1})}\leftrightarrow\alpha_{2n-i}^{(A_{2n-1})} $. A new lattice obtained by taking the $g$-invariant part of $\Lambda_R(A_{2n-1})$ is close to the root lattice of $C_n$:
\begin{equation}
\begin{aligned}
        \Lambda_R^g(A_{2n-1}):=&\{ p\in \Lambda_R(A_{2n-1})| g(p)=p\}\\
=&\Z\qty(\alpha_1^{(A_{2n-1)}}+\alpha_{2n-1}^{(A_{2n-1)}})\oplus\cdots\oplus\Z\qty(\alpha_{n-1}^{(A_{2n-1)}}+\alpha_{n+1}^{(A_{2n-1)}})\oplus\Z\alpha_n^{(A_{2n-1})}\\
        \cong&\sqrt{2}\qty(\Lambda_R(C_n)+\frac{1}{2}\Z\alpha_n^{(C_n)}),
\end{aligned}
\end{equation}
where $\cong $ means an isomorphism of the abelian group, and here we identified $C_n$ roots from $A_{2n-1}$ roots as follows:
\begin{equation}
    \begin{aligned}
        &\alpha_i^{(C_n)}=\frac{1}{\sqrt{2}}\qty(\alpha_i^{(A_{2n-1})}+\alpha_{2n-i}^{(A_{2n-1})}),\quad\text{for}\quad1\leq i\leq n-1,\\
&\alpha_n^{(C_n)}=\sqrt{2}\alpha_n^{(A_{2n-1})}.\\
    \end{aligned}
\end{equation}
This process is called \textit{folding}, since it corresponds to the folding of the Dynkin diagram (see Figure \ref{Fig_AtoC}).

\subsubsection{\texorpdfstring{$D_4\to B_3$}{D4->B3}}
Another example is folding  $D_4$ into $B_3$.
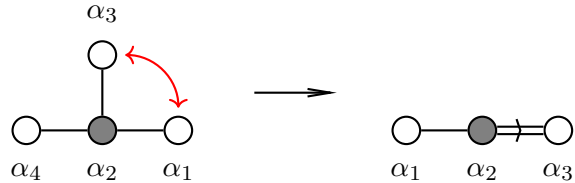
\begin{figure}[htb]
\centering
\begin{tikzpicture}[scale=1.0]

\node[thick, circle, draw, fill=white] (1) at (0,0) {};
\node[thick, circle, draw, fill=gray] (2) at (1,0) {};
\node[thick, circle, draw, fill=white] (3) at (2,0) {};
\node[thick, circle, draw, fill=white] (4) at (1,1) {};

\draw[thick] (1) -- (2);
\draw[thick] (2) -- (3);
\draw[thick] (2) -- (4);

\node[below=0.3cm] at (1) {$\alpha_4$};
\node[below=0.3cm] at (2) {$\alpha_2$};
\node[below=0.3cm] at (3) {$\alpha_1$};
\node[above=0.3cm] at (4) {$\alpha_3$};

\node[thick, circle, draw, fill=white] (1) at (5,0) {};
\node[thick, circle, draw, fill=gray] (2) at (6,0) {};
\node[thick, circle, draw, fill=white] (3) at (7,0) {};

\draw[thick] (1) -- (2);
\draw[thick] (3,0.5)--(4,0.5);
\draw[thick] (3.7,0.55)--(4,0.5)--(3.7,0.45);
\draw[thick, red, <->] (1.3,1) to [out=0,in=90] (2,0.3);

\draw[double distance=2pt, thick] (2) -- (3);
\draw[thick] (6.45,0.15) -- (6.5,0) -- (6.45,-0.15);

\node[below=0.3cm] at (1) {$\alpha_1$};
\node[below=0.3cm] at (2) {$\alpha_2$};
\node[below=0.3cm] at (3) {$\alpha_3$};

\end{tikzpicture}
\caption{Folding  $D_4$ into $B_3$}
\end{figure}

The root lattice of $D_4$ admits a $\Z_2$ symmetry $g$ which flips $\alpha_1^{(D_4)}\leftrightarrow\alpha_3^{(D_4)}$. Then we identified $B_3$ roots from $D_4$ roots  as follows:

\begin{equation}
    \begin{aligned}
        \alpha_1^{(B_3)}=&\alpha_4^{(D_4)},\\
        \alpha_2^{(B_3)}=&\alpha_2^{(D_4)},\\
        \alpha_3^{(B_3)}=&\frac{1}{2}\qty(\alpha_1^{(D_4)}+\alpha_3^{(D_4)}).\\
    \end{aligned}
\end{equation}

The invariant sublattice of $D_4$ root lattice under $g$ is 

\begin{equation}
    \begin{aligned}
&\Z\alpha^{(D_4)}_4\oplus\Z\alpha^{(D_4)}_2\oplus\Z(\alpha^{(D_4)}_1+\alpha^{(D_4)}_3)\\
=&\Z\alpha^{(B_3)}_1\oplus\Z\alpha^{(B_3)}_2\oplus\Z2\alpha_3^{(B_3)}.
    \end{aligned}
\end{equation}

\subsubsection{\texorpdfstring{$E_6\to F_4$}{E6->F4}}

The last one used in this paper is folding $E_6$ into $F_4$.
\begin{figure}[h]
    \begin{center}
       \begin{tikzpicture}[scale=1.0]
            \node[thick, circle, draw, fill=white] (1) at (2,2) {};
            \node[thick, circle, draw, fill=white] (3) at (2,1) {};
            \node[thick, circle, draw, fill=gray] (4) at (2,0) {};
            \node[thick, circle, draw, fill=white] (5) at (3,0) {};
            \node[thick, circle, draw, fill=white] (6) at (4,0) {};
            \node[thick, circle, draw, fill=white] (2) at (1,0) {};

            \draw[thick] (1) -- (3);
            \draw[thick] (3) -- (4);
            \draw[thick] (4) -- (5);
            \draw[thick] (5) -- (6);
            \draw[thick] (4) -- (2); 
            \draw[thick, red, <->] (2.4,1.5) to [out=0,in=90] (3.7,0.4);

            \node[left=0.3cm] at (1) {$\alpha_1$};
            \node[left=0.3cm] at (3) {$\alpha_3$};
            \node[below=0.3cm] at (4) {$\alpha_4$};
            \node[below=0.3cm] at (5) {$\alpha_5$};
            \node[below=0.3cm] at (6) {$\alpha_6$};
            \node[below=0.3cm] at (2) {$\alpha_2$};
      
            \node[thick, circle, draw, fill=white] (1) at (7,0) {};
            \node[thick, circle, draw, fill=gray] (2) at (8,0) {};
            \node[thick, circle, draw, fill=white] (3) at (9,0) {};
            \node[thick, circle, draw, fill=white] (4) at (10,0) {};
            
            \draw[thick] (1) -- (2);
            \draw[double distance=2pt, thick] (2) -- (3); 
            \draw[thick] (3) -- (4);
            \draw[thick] (5,0.5)--(6,0.5);
\draw[thick] (5.7,0.55)--(6,0.5)--(5.7,0.45);

            \draw[thick] (8.55,0.15) -- (8.5,0) -- (8.55,-0.15);

            \node[below=0.3cm] at (1) {$\alpha_4$};
            \node[below=0.3cm] at (2) {$\alpha_3$};
            \node[below=0.3cm] at (3) {$\alpha_2$};
            \node[below=0.3cm] at (4) {$\alpha_1$};
        \end{tikzpicture}
        \caption{Folding  $E_6$ into $F_4$}
        \label{Fig_B4_Dynkin}
    \end{center}
\end{figure}
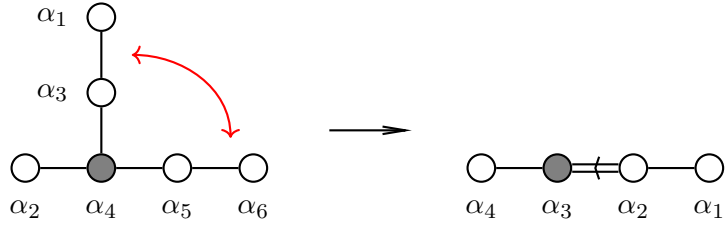

The root lattice of $E_6$ admits a $\Z_2$ symmetry $g$ which flips $\alpha_1^{(E_6)}\leftrightarrow\alpha_6^{(E_6)}$ and $\alpha_3^{(E_6)}\leftrightarrow\alpha_5^{(E_6)}$ simultaneously. Then we identified $F_4$ roots from $E_6$ roots  as follows:

\begin{equation}
    \begin{aligned}
        \alpha_1^{(F_4)}=&\frac{1}{\sqrt{2}}\qty(\alpha_1^{(E_6)}+\alpha_6^{(E_6)}),\\
        \alpha_2^{(F_4)}=&\frac{1}{\sqrt{2}}\qty(\alpha_3^{(E_6)}+\alpha_5^{(E_6)}),\\
        \alpha_3^{(F_4)}=&\frac{1}{\sqrt{2}}\alpha_4^{(E_6)},\\
        \alpha_4^{(F_4)}=&\frac{1}{\sqrt{2}}\alpha_2^{(E_6)}.\\
    \end{aligned}
\end{equation}

The invariant sublattice of $E_6$ root lattice under $g$ is 

\begin{equation}
    \begin{aligned}
&\Z\qty(\alpha_1^{(E_6)}+\alpha_6^{(E_6)})\oplus\Z\qty(\alpha_3^{(E_6)}+\alpha_5^{(E_6)})\oplus\Z\alpha_4^{(E_6)}\oplus\Z\alpha_2^{(E_6)}\\
=&\sqrt{2}\alpha_1^{(F_4)}\oplus\sqrt{2}\Z\alpha_2^{(F_4)}\oplus\frac{1}{\sqrt{2}}\Z\alpha_3^{(F_4)}\oplus\frac{1}{\sqrt{2}}\Z\alpha_1^{(F_4)}.
    \end{aligned}
\end{equation}

\section{Review of 9d Results}\label{sec:9d_review}
In this section, we review the 9d theories obtained in \cite{Hamada:2025JHEP}. The results are summarized in Table \ref{tab:9d_summary}. 
The massless states in the untwisted sectors consist of the graviton, $B$-field, dilaton, gauge fields, gauginos, and the adjoint scalars. The spectrum of the twisted sectors can be read off from \eqref{eq:twisted_partition} by $q,\bar{q}$-expansions. Due to the level matching condition, we only have to study terms in which the powers of $q$ and $\bar{q}$ are matched.

\begin{table}[h]
    \centering
    \begin{tabular}{|c|c|c|c|c|}
  \hline
  \textbf{SUSY}
  &\cancel{\textbf{SUSY}}
  & \multicolumn{2}{c|}{\textbf{Twisted sector}} \\ \cline{3-4}
  theory& theory
  & $\bar{O}_8$ & $\bar{C}_8$ \\ \hline
  $A_{17}$ & 
  $C_9$  
  & $\mathbf{1}$ & $\mathbf{152}$\\ \hline
  $A_{15}+2A_1$ 
  & $C_8 + A_1$ 
  & $\mathbf{1}$ & $
(\mathbf{119},\mathbf{1})\oplus(\mathbf{1},\mathbf{3})$\\ \hline
  $A_{11}+2A_2+2A_1$ 
  & $C_6 + A_2 + A_1$ & $\mathbf{1}$ &  $\qty(\mathbf{65},\mathbf{1},\mathbf{1})\oplus\qty(\mathbf{1},\mathbf{8},\mathbf{1})\oplus\qty(\mathbf{1},\mathbf{1},\mathbf{3})$ 
 \\ \hline
  $A_9+2A_4$ 
  &$C_5+A_4$ & $\mathbf{1}$ &    $ (\mathbf{44},\mathbf{1})\oplus(\mathbf{1},\mathbf{24})$
\\ \hline
  $A_7+2D_5$ 
  &$C_4+D_5$ & $\mathbf{1}$ &
$\qty(\mathbf{27},\mathbf{1})\oplus\qty(\mathbf{1},\mathbf{45})$  \\ \hline
  $A_5+2E_6$ 
  &$C_3+E_6$ & $\mathbf{1}$ & $\qty(\mathbf{14},\mathbf{1})\oplus\qty(\mathbf{1},\mathbf{78})$
\\ \hline
  $A_3+2E_7$ 
  & $C_2+E_7$ 
  & $\mathbf{1}$ & $\qty(\mathbf{5},\mathbf{1})\oplus\qty(\mathbf{1},\mathbf{133})$\\ \hline
  $A_1+2E_8$ 
  & $C_1+E_8$ 
  & $\mathbf{1}$ &    $\qty(\mathbf{1},\mathbf{248})$
 \\ \hline
\end{tabular}
\caption{The list of 9d non-supersymmetric heterotic strings obtained from 9d supersymmetric heterotic strings by orbifolding the outer automorphism and the spacetime fermion parity.}
\label{tab:9d_summary}
\end{table}

We note that as discussed in \cite{Font:2020rsk}, the gauge groups for the 9d SUSY theories can be obtained by removing two appropriate nodes from the generalized Dynkin diagram shown in Figure \ref{Fig_9dsusy}. Similarly, the gauge groups for the 9d \cancel{SUSY} theories can be obtained by removing one appropriate node from the generalized Dynkin diagram shown in Figure \ref{Fig_9dnonsusy}.  After investigating the maximal enhancements of the $E_8$ string theory on $S^1$ in subsection \ref{sec:9d_maximal}, we propose the generalized Dynkin diagram for the 9d SUSY theories which can be used for obtaining the allowed gauge groups. The rules for the 9d generalized Dynkin diagram is explained in subsection \ref{sec:9d_Dynkin}.

\begin{figure}[htb]
    \begin{center}
        \begin{tikzpicture}[scale=0.8]
            \tikzset{double circle/.style={draw, circle, double, thick, minimum size=10pt, inner sep=1pt}}
            
    \node[thick, circle, draw, fill=white] (1) at (0,0) {};
    \node[thick, circle, draw, fill=white] (2) at (1,0) {};
    \node[thick, circle, draw, fill=white] (3) at (2,0) {};
    \node[thick, circle, draw, fill=white] (4) at (3,0) {};
    \node[thick, circle, draw, fill=white] (5) at (4,0) {};
    \node[thick, circle, draw, fill=white] (6) at (5,0) {};
    \node[thick, circle, draw, fill=white] (7) at (1,1) {};
    \node[thick, circle, draw, fill=white] (8) at (1,2) {};
    \node[thick, circle, draw, fill=white] (0) at (6,0) {};
    \node[thick, circle, draw, fill=gray] (C1) at (7,0) {};

        \node[thick, circle, draw, fill=white] (1') at (14,0) {};
    \node[thick, circle, draw, fill=white] (2') at (13,0) {};
    \node[thick, circle, draw, fill=white] (3') at (12,0) {};
    \node[thick, circle, draw, fill=white] (4') at (11,0) {};
    \node[thick, circle, draw, fill=white] (5') at (10,0) {};
    \node[thick, circle, draw, fill=white] (6') at (9,0) {};
    \node[thick, circle, draw, fill=white] (7') at (13,1) {};
    \node[thick, circle, draw, fill=white] (8') at (13,2) {};
    \node[thick, circle, draw, fill=white] (0') at (8,0) {};

    \draw[thick] (1) -- (2);
    \draw[thick] (2) -- (3);
    \draw[thick] (3) -- (4);
    \draw[thick] (4) -- (5);
    \draw[thick] (5) -- (6);
    \draw[thick] (2) -- (7);
    \draw[thick] (7) -- (8);
    \draw[thick] (6) -- (0);
    \draw[thick] (0) -- (C1);
    \draw[thick] (C1) -- (0');
\draw[thick] (1') -- (2');
    \draw[thick] (2') -- (3');
    \draw[thick] (3') -- (4');
    \draw[thick] (4') -- (5');
    \draw[thick] (5') -- (6');
    \draw[thick] (2') -- (7');
    \draw[thick] (7') -- (8');
    \draw[thick] (6') -- (0');

    \node[below=0.3cm] at (1) {};
    \node[below=0.3cm] at (2) {};
    \node[below=0.3cm] at (3) {};
    \node[below=0.3cm] at (4) {};
    \node[below=0.3cm] at (5) {};
    \node[below=0.3cm] at (6) {};
    \node[left=0.3cm] at (7) {};
    \node[left=0.3cm] at (8) {};
    \node[below=0.3cm] at (0) {};
    \node[below=0.3cm] at (C1) {};
    \node[below=0.3cm] at (1') {};
    \node[below=0.3cm] at (2') {};
    \node[below=0.3cm] at (3') {};
    \node[below=0.3cm] at (4') {};
    \node[below=0.3cm] at (5') {};
    \node[below=0.3cm] at (6') {};
    \node[right=0.3cm] at (7') {};
    \node[right=0.3cm] at (8') {};
    \node[below=0.3cm] at (0') {};
        \end{tikzpicture}
        \caption{Generalized Dynkin diagram for 9d SUSY theories.}\label{Fig_9dsusy}
    \end{center}
\end{figure}

\begin{figure}[htb]
    \begin{center}
        \begin{tikzpicture}[scale=1.0]
            \tikzset{double circle/.style={draw, circle, double, thick, minimum size=10pt, inner sep=1pt}}
            
    \node[thick, circle, draw, fill=white] (1) at (0,0) {};
    \node[thick, circle, draw, fill=white] (2) at (1,0) {};
    \node[thick, circle, draw, fill=white] (3) at (2,0) {};
    \node[thick, circle, draw, fill=white] (4) at (3,0) {};
    \node[thick, circle, draw, fill=white] (5) at (4,0) {};
    \node[thick, circle, draw, fill=white] (6) at (5,0) {};
    \node[thick, circle, draw, fill=white] (7) at (1,1) {};
    \node[thick, circle, draw, fill=white] (8) at (1,2) {};
    \node[thick, circle, draw, fill=white] (0) at (6,0) {};
    \node[thick, circle, draw, fill=gray] (C1) at (7,0) {};

    \draw[thick] (1) -- (2);
    \draw[thick] (2) -- (3);
    \draw[thick] (3) -- (4);
    \draw[thick] (4) -- (5);
    \draw[thick] (5) -- (6);
    \draw[thick] (2) -- (7);
    \draw[thick] (7) -- (8);
    \draw[thick] (6) -- (0);
    \draw[double distance=2pt, thick] (0) -- (C1);
    \draw[thick] (6.6,0.15) -- (6.4,0.0) -- (6.6,-0.15);

        \end{tikzpicture}
        \caption{Generalized Dynkin diagram for 9d \cancel{SUSY} theories.}\label{Fig_9dnonsusy}
    \end{center}
\end{figure}

\section{Gauging lattice CFTs by folding Dynkin diagrams in 8d}\label{sec:8d_folding}
In this section, we perform the orbifolding from the 8d SUSY theories whose charge lattices have a $\mathbb{Z}_2$ symmetry, and study the massless spectrum. 
We obtain 22 gauge symmetries, as is summarized in Table \ref{tab:8d_summary}.
Here the number $\#$ indicates the one in Table 12 of \cite{Font:2020rsk}. 

In the previous work \cite{Hamada:2025JHEP}, the eight non-SUSY theories with the 9d gauge groups plus a $C_1$ factor were obtained from the 8d SUSY theories whose gauge groups are expressed as the 9d possible gauge groups and an extra $A_1$. We skip the calculations for these cases in this paper, but we include them into Table \ref{tab:8d_summary}. These eight cases correspond to $\#57,81,106,111,121,222,257,296$.

\begin{table}[h]
    \centering
    \scalebox{0.9}{
    \begin{tabular}{|c||c|c|c|c|} \hline
        $\#$ & \textbf{SUSY} 
        &\cancel{\textbf{SUSY}} 
        & \multicolumn{2}{c|}{\textbf{Twisted sector}} \\ \cline{4-5}
        & theory & theory 
        & $\bar{O}_8$ & $\bar{C}_8$ \\ \hline
        $1$ & $6A_3$ & $2C_2+2A_3$ & $\mathbf{1}$ & $((\mbf{5},\mbf{1})\oplus(\mbf{1},\mbf{5}),\mbf{1},\mbf{1}), (\mbf{1},\mbf{1},(\mbf{15},\mbf{1})\oplus(\mbf{1},\mbf{15}))$ \\ \hline
        $2$ & $2A_1+4A_4$ & $2C_1+2A_4$ & $\mathbf{1}$ & $(\mbf{1},\mbf{1},(\mbf{24},\mbf{1})\oplus(\mbf{1},\mbf{24}))$ \\ \hline
        $5$ & $4A_2 + 2A_5$ & $2A_2+2C_3$ & $\mathbf{1}$ & $((\mbf{8},\mbf{1})\oplus(\mbf{1},\mbf{8}),\mbf{1},\mbf{1}), (\mbf{1},\mbf{1},(\mbf{14},\mbf{1})\oplus(\mbf{1},\mbf{14}))$ \\ \hline
        $6$ & $A_3 + 3A_5$ & $C_2 + C_3 + A_5$ & $\mathbf{1}$ & $\qty(\mathbf{5},\mathbf{1},\mathbf{1}),~ \qty(\mathbf{1},\mathbf{14},\mathbf{1}), ~ \qty(\mathbf{1},\mathbf{1},\mathbf{35})$ \\ \hline
        $25$ & $4A_1 + 2A_7$ & $2A_1 + 2C_4$ & $\mathbf{1}$ & $((\mbf{3},\mbf{1})\oplus(\mbf{1},\mbf{3}),\mbf{1},\mbf{1}), (\mbf{1},\mbf{1},(\mbf{27},\mbf{1})\oplus(\mbf{1},\mbf{27}))$ \\ \hline
        $28$ & $2A_1 + 3A_3 + A_7$ & $A_1 + C_2 + A_3 + C_4$ & $\mathbf{1}$ & $(\mbf{1},\mbf{5},\mbf{1},\mbf{1}),~(\mbf{1},\mbf{1},\mbf{15},\mbf{1}),~(\mbf{1},\mbf{1},\mbf{1},\mbf{27})~$ \\ \hline
        $54$ & $2A_9$ & $2C_5$ & $\mathbf{1}$ & $\qty(\mbf{44},\mbf{1}),~ \qty(\mbf{1},\mbf{44})$ \\ \hline
        $57$ & $A_1 + 2A_4 + A_9$ & $C_1 + A_4 + C_5$ & $\mathbf{1}$ & $(\mbf{1},\mbf{24},\mbf{1}),(\mbf{1},\mbf{1},\mbf{44})$ \\ \hline
        $81$ & $3A_1 + 2A_2 + A_{11}$ & $A_1 + C_1 + A_2 + C_6$ & $\mathbf{1}$ & $(\mbf{1},\mbf{3},\mbf{1},\mbf{1}),(\mbf{1},\mbf{1},\mbf{8},\mbf{1}),(\mbf{1},\mbf{1},\mbf{1},\mbf{65})$ \\ \hline
        $83$ & $2A_2 + A_3 + A_{11}$ & $A_2 + C_2 + C_6$ & $\mathbf{1}$ & $(\mbf{8},\mbf{1},\mbf{1}),~(\mbf{1},\mbf{5},\mbf{1}),~(\mbf{1},\mbf{1},\mbf{65})$ \\ \hline
        $87$ & $2A_1 + A_5 + A_{11}$ & $A_1 + C_3 + C_6$ & $\mathbf{1}$ & $(\mbf{3},\mbf{1},\mbf{1}),~(\mbf{1},\mbf{14},\mbf{1}),~(\mbf{1},\mbf{1},\mbf{65})$ \\ \hline
        $106$ & $3A_1 + A_{15}$ & $A_1 + C_1 + C_8$ & $\mathbf{1}$ & $(\mbf{3},\mbf{1},\mbf{1}),(\mbf{1},\mbf{1},\mbf{119})$ \\ \hline
        $108$ & $A_3 + A_{15}$ & $C_2 + C_8$ & $\mathbf{1}$ & $\qty(\mathbf{5},\mathbf{1}),~\qty(\mathbf{1},\mathbf{119})$ \\ \hline
        $111$ & $A_1 + A_{17}$ & $C_1 + C_9$ & $\mathbf{1}$ & $(\mbf{1},\mbf{152})$ \\ \hline
        $121$ & $A_1 + A_7 + 2D_5$ & $C_1 + C_4 +D_5$ & $\mathbf{1}$ & $(\mbf{1},\mbf{27},\mbf{1}),(\mbf{1},\mbf{1},\mbf{44})$ \\ \hline
        $137$ & $2A_3 + 2D_6$ & $2C_2 + D_6$ & $\mathbf{1}$ & $((\mbf{5},\mbf{1})\oplus(\mbf{1},\mbf{5}),\mbf{1}),~(\mbf{1},\mbf{1},\mbf{66})$ \\ \hline
        $219$ & $3E_6$ & $E_6 + F_4$ & $\mathbf{1}$ & $(\mbf{78},\mbf{1}),\qty(\mbf{1},\mbf{26})$ \\ \hline
        $222$ & $A_1 + A_5 + 2E_6$ & $C_1 + C_3 + E_6$ & $\mathbf{1}$ & $(\mbf{1},\mbf{14},\mbf{1}),~(\mbf{1},\mbf{1},\mbf{78})$ \\ \hline
        $257$ & $A_1 + A_3 + 2E_7$ & $C_1 + C_2 + E_7$ & $\mathbf{1}$ & $(\mbf{1},\mbf{5},\mbf{1}),~(\mbf{1},\mbf{1},\mbf{133})$ \\ \hline
        $279$ & $D_4 + 2E_7$ & $B_3 + E_7$ & $\mathbf{1}$ & $(\mbf{7},\mbf{1}),\qty(\mbf{1},\mbf{133})$ \\ \hline
        $296$ & $2A_1 + 2E_8$ & $2C_1 + E_8$ & $\mathbf{1}$ & $(\mbf{1},\mbf{248})$ \\ \hline
        $297$ & $A_2 + 2E_8$ & $A_2+E_8$ & $\mathbf{1}$ & $(\mbf{1},\mbf{248})$ \\ \hline
    \end{tabular}
    }
    \caption{The list of 8d (non-)supersymmetric heterotic strings and lighter particles in the twisted sector.  $\bar{O}_8$ is the singlet representation of spacetime $\SO(8)$, and $\bar{C}_8$ is the conjugacy spinor representation. $\mathbf{1}$ in $\bar{O}_8$ is a tachyon, and representations in $\bar{C}_8$ are fermions. }
    \label{tab:8d_summary}
\end{table}

\subsection{\texorpdfstring{$6A_3\to 2C_2+2A_3~(\#1)$}{6A3 -> 2C2+2A3}}

We start from the even self-dual lattice:
\begin{equation}
    \begin{aligned}
\Gamma^{(6A_3)}_{18,2}=&\Lambda_R(6A_3)+\Z\qty(0,2,1,1,1,1)\omega_1^{(A_3)}+\Z\qty(1,1,0,2,1,-1)\omega_1^{(A_3)}
+\Z(0;2,0)+\Z\qty(0;0,2)\\
+&\Z\qty(\qty(0,1,1,0,0,1)\omega_1^{(A_3)};\frac{1}{2},0)
+\Z\qty(\qty(1,0,0,-1,1,0)\omega_1^{(A_3)};0,\frac{1}{2}).
    \end{aligned}
\end{equation}
This lattice admits the following symmetry:
\begin{equation}
\begin{aligned}
g:&\qty(x_1^{(A_3)},x_2^{(A_3)},x_3^{(A_3)},x_4^{(A_3)},x_5^{(A_3)},x_6^{(A_3)})\\
\mapsto& \qty(\tilde{x}_1^{(A_3)},\tilde{x}_2^{(A_3)},-x_4^{(A_3)},-x_3^{(A_3)},-x_6^{(A_3)},-x_5^{(A_3)}),\\
\end{aligned}
\end{equation}
where $\tilde{x}^{(A_3)}$ is obtained from exchanging $\alpha_i$ and $\alpha_{2n-i}$ in $x^{(A_3)}$ for all $i=1,\cdots,2n-1$.

The invariant lattice and its dual lattice are
\begin{equation}
    \begin{aligned}
        I=&\sqrt{2}\qty(\Lambda_R(2C_2+A_3)+\frac{1}{2}\Z\alpha_2^{(C_2)}+\frac{1}{2}\Z\alpha_2^{(C_2)}+\Z(0;\sqrt{2},0)+\Z\qty(0;0,\sqrt{2}))\\
        +&\sqrt{2}\Z\qty(\qty(0,0),(2,-2)\omega_1^{(A_3)})+\sqrt{2}\Z\qty(\qty(\frac{1}{2},\frac{1}{2})\omega_2^{(C_2)},(0,2)\omega_1^{(A_3)})\\
        +&\sqrt{2}\Z\qty((0,0),(1,-1)\omega_1^{(A_3)};\frac{1}{\sqrt{2}},0)+\sqrt{2}\Z\qty((0,0),(1,1)\omega_1^{(A_3)};0,\frac{1}{\sqrt{2}})\Biggr),\\
        I^\ast=&I+\frac{1}{\sqrt{2}}\Lambda_R(2C_2+2A_3).
    \end{aligned}
\end{equation}

By gathering the eight neutral elements and elements $(p_L,p_R)\in I^\ast$ that satisfy $p_L^2=1, p_R^2=0$, we obtain the following representation of $2C_2+2A_3$  as massless fermions:
\begin{equation}
((\mbf{5},\mbf{1})\oplus(\mbf{1},\mbf{5}),\mbf{1},\mbf{1}), (\mbf{1},\mbf{1},(\mbf{15},\mbf{1})\oplus(\mbf{1},\mbf{15})).
\end{equation}

The gauge symmetry is given by

\begin{equation}
    \Sp(2)\times \Sp(2)\times \frac{\SU(4)\times\SU(4)\times \U(1)\times \U(1)}{\Z_2\times\Z_2}.
\end{equation}

\subsection{\texorpdfstring{$2A_1+4A_4\to 2C_1+2A_4~(\#2)$}{2A1+4A4-> 2C1+2A4}}
The even self-dual lattice is:
\begin{equation}
\begin{aligned}
\Gamma^{(2A_1+4A_4)}=&\Lambda_R(2A_1+4A_4)+\Z\qty(0,0,(1,1,2,2)\omega_1^{(A_4)})+\Z(0;\sqrt{10},0)+\Z(0;\sqrt{10},0)\\
    +&\Z\qty(\omega^{(A_1)}_1,0,(1,-1,0,0)\omega^{(A_4)}_1;\frac{1}{\sqrt{10}},0)
    +\Z\qty(0,\omega^{(A_1)}_1,(0,0,1,-1)\omega_1^{(A_4)};0,\frac{1}{\sqrt{10}}).
\end{aligned}
\end{equation}
This lattice admits the following symmetry:
\begin{equation}
\begin{aligned}
    g: &\qty(x^{(A_1)}_1,x_2^{(A_1)},x_1^{(A_4)},x_2^{(A_4)},x_3^{(A_4)},x_4^{(A_4)})\\
    \mapsto &\qty(x_1^{(A_1)},x_2^{(A_1)},-x_2^{(A_4)},-x_1^{(A_4)},-x_4^{(A_4)},-x_3^{(A_4)}).
\end{aligned}
\end{equation}
The invariant lattice and its dual lattice are
\begin{equation}
    \begin{aligned}
        I=&\sqrt{2}\qty(\Lambda_R\left(2 C_1+2 A_4\right)+\mathbb{Z} \frac{1}{2} \alpha_1^{(C_1)}+\Z \frac{1}{2} \alpha_1^{(C_1)}+\Z(0;\sqrt{5} ,0)+\Z(0;0,\sqrt{5}))\\
 \quad+&\sqrt{2} \Z\qty(\qty(\frac{1}{2},0)\omega_1^{(C_1)},(1,0)\omega_1^{(A_4)};\frac{1}{\sqrt{20}}, 0)+\sqrt{2} \Z\qty(\qty(0, \frac{1}{2} )\omega_1^{(C_1)}, (0, 1)\omega_1^{\left(A_4\right)};0, \frac{1}{\sqrt{20}}),\\
 I^\ast=&I+\frac{1}{\sqrt{2}}\Lambda_R(2 C_1+2 A_4).
    \end{aligned}
\end{equation}
By gathering the eight neutral elements and the elements $(p_L,p_R)\in I^\ast$ that satisfy $p_L^2=1, p_R^2=0$, we obtain the following representation of $2C_1+2A_4$  as massless fermions:
\begin{equation}
(\mbf{1},\mbf{1},(\mbf{24},\mbf{1})\oplus(\mbf{1},\mbf{24})).
\end{equation}

The gauge symmetry is 

\begin{equation}
    \frac{\Sp(1)\times\Sp(1)\times \SU(5)\times\SU(5)\times \U(1)\times \U(1)}{\Z_{10}\times\Z_{10}}.
\end{equation}

\subsection{\texorpdfstring{$4A_2+2A_5\to 2A_2+2C_3~(\#5)$}{4A2+2A5 -> 2A2+2C3}}
The even self-dual lattice is:
\begin{equation}
    \begin{aligned}
\Gamma^{(4A_2+2A_5)}_{18,2}=&\Lambda_R(4A_2+2A_5)+\Z\qty((0,0,1,1)\omega_1^{(A_2)},(2,2)\omega_1^{(A_5)})+Z\qty((1,1,0,0)\omega_1^{(A_2)},(2,-2)\omega_1^{(A_5)})\\
+&\Z\qty((1,0,0,2)\omega_1^{(A_2)},(0,1)\omega_1^{(A_5)};\frac{1}{\sqrt{6}},0)+\Z\qty((0,1,2,0)\omega_1^{(A_2)},(1,0)\omega_1^{(A_5)};0,\frac{1}{\sqrt{6}})\\
+&\Z(0;\sqrt{6},0)+\Z(0;0,\sqrt{6}).
    \end{aligned}
\end{equation}
This lattice admits the following symmetry:
\begin{equation}
\begin{aligned} g:&\qty(x_1^{(A_2)},x_2^{(A_2)},x_3^{(A_2)},x_4^{(A_2)},x_1^{(A_5)},x_2^{(A_5)})\\
\mapsto&\qty(-x_2^{(A_2)},-x_1^{(A_2)},-x_4^{(A_2)},-x_3^{(A_2)},\tilde{x}_1^{(A_5)},\tilde{x}_2^{(A_5)}).
\end{aligned}
\end{equation}
The invariant lattice and its dual lattice are
\begin{equation}
    \begin{aligned}
        I=&\sqrt{2}\qty(\Lambda_R(2A_2+2C_3)+\frac{1}{2}\Z\alpha_3^{(C_3)}+\frac{1}{2}\Z\alpha_3^{(C_3)}+\Z(0;\sqrt{3},0)+\Z(0;0,\sqrt{3}))\\
        +&\sqrt{2}\Z\qty(\qty(1,1)\omega_1^{(A_2)},(0,0);\frac{1}{\sqrt{3}},0)+\sqrt{2}\Z\qty((-1,-1)\omega_1^{(A_2)},(0,0);0,\frac{1}{\sqrt{3}})\\
        +&\sqrt{2}\Z\qty((0,0),\qty(0,\frac{1}{2}\omega_3^{(C_3)});\frac{\sqrt{3}}{2},0)+\sqrt{2}\Z\qty((0,0),\qty(\frac{1}{2}\omega_3^{(C_3)},0);0,\frac{\sqrt{3}}{2}),\\
        I^\ast=&I+\frac{1}{\sqrt{2}}\Lambda_R(2A_2+2C_3).
    \end{aligned}
\end{equation}
By gathering the eight neutral elements and the elements $(p_L,p_R)\in I^\ast$ that satisfy $p_L^2=1, p_R^2=0$, we obtain the following representation of $2A_2+2C_3$  as massless fermions:
\begin{equation}
((\mbf{8},\mbf{1})\oplus(\mbf{1},\mbf{8}),\mbf{1},\mbf{1}), (\mbf{1},\mbf{1},(\mbf{14},\mbf{1})\oplus(\mbf{1},\mbf{14})).
\end{equation}

The gauge group is 

\begin{equation}
    \frac{\Sp(3)\times \Sp(3)\times \SU(3)\times \SU(3)\times \U(1)\times \U(1)}{\Z_3\times \Z_3\times \Z_2\times \Z_2}.
\end{equation}

\subsection{\texorpdfstring{$A_3+3A_5\to C_2+C_3+A_5~(\#6)$}{A3+3A5->C2+C3+A5(\#6)}}
We start from the even self-dual lattice:
\begin{equation}
    \begin{aligned}
\Gamma^{(A_3+3A_5)}_{18,2}=&\Lambda_R(A_3+3A_5)+\Z(0;2,0)+\Z(0;0,\sqrt{6})+\Z\qty(2\omega_1^{(A_3)},(2,5,5)\omega_1^{(A_5)};0,0)\\
    +&\Z\qty(\omega_1^{(A_3)},(0,3,0)\omega_1^{(A_5)};\frac{1}{2},0)+\Z\qty(0,(1,2,0)\omega_1^{(A_5)};0,\frac{1}{\sqrt{6}}).
    \end{aligned}
\end{equation}
This lattice has the following symmetry:
\begin{equation}
    \begin{aligned}
g:&\qty(x^{(A_3)},x^{(A_5)},x^{(A_5)}_1,x^{(A_5)}_2;x^{(R)})\\
    \mapsto&\qty(\tilde{x}^{(A_3)},\tilde{x}^{(A_5)},-x^{(A_5)}_2,-x^{(A_5)}_1;x^{(R)}).
    \end{aligned}
\end{equation}
The invariant lattice $I$ and its dual lattice  $I^\ast$ are
\begin{equation}
    \begin{aligned}
I=&\sqrt{2}\qty(\Lambda_R(C_2+C_3+A_5)+\Z(0;\sqrt{2},0)+\Z(0;0,\sqrt{3})+\frac{1}{2}\Z\alpha_2^{(C_2)}+\frac{1}{2}\Z\alpha_3^{(C_3)})\\
+&\sqrt{2}\Z\qty(\frac{1}{2}\omega_2^{(C_2)},0,3\omega_1^{(A_5)};0,0) 
    +\sqrt{2}\Z\qty(\frac{1}{2}\omega_2^{(C_2)},0,0;\frac{1}{\sqrt{2}},0)+\sqrt{2}\Z\qty(0,0,2\omega_1^{(A_5)};0,\frac{1}{\sqrt{3}})\\
    +&\sqrt{2}\Z\qty(0,\frac{1}{2}\omega_3^{(C_3)},\omega_1^{(A_5)};\frac{1}{\sqrt{2}},\frac{1}{2\sqrt{3}}),\\
    I^\ast=&I+\frac{1}{\sqrt{2}}\Lambda_R(C_2+C_3+A_5).
    \end{aligned}
\end{equation}
By gathering the eight neutral elements and the elements $(p_L,p_R)\in I^\ast$ that satisfy $p_L^2=1, p_R^2=0$, we obtain the following representation of $C_2+C_3+A_5$  as massless fermions:
\begin{equation}
    \qty(\mathbf{5},\mathbf{1},\mathbf{1}),~ \qty(\mathbf{1},\mathbf{14},\mathbf{1}), ~ \qty(\mathbf{1},\mathbf{1},\mathbf{35}).
\end{equation}

The gauge group is

\begin{equation}
    \Sp(2)\times \frac{\Sp(3)\times\SU(6)\times  \U(1)\times \U(1)}{\Z_6}.
\end{equation}

\subsection{\texorpdfstring{$2A_7+4A_1\to 2C_4+2A_1~(\#25)$}{2A7+4A1-> 2C4+2A1}}
The even self-dual lattice is:
\begin{equation}
    \begin{aligned}
\Gamma^{(2A_7+4A_1)}_{18,2}=&\Lambda_R(2A_7+4A_1)+\Z(0;2,0)+\Z(0;0,2)\\
+&\Z\qty((2,2)\omega_1^{(A_7)},(1,1)\omega_1^{(A_1)},(0,0))+\Z\qty((4,0)\omega_1^{(A_7)},(1,1)\omega_1^{(A_1)},(1,1)\omega_1^{(A_1)})\\
+&\Z\qty((1,1)\omega_1^{(A_7)},(0,1)\omega_1^{(A_1)},(0,0);\frac{1}{2},0)+\Z\qty((1,-1)\omega_1^{(A_7)},(0,0),(0,1)\omega_1^{(A_1)};0,\frac{1}{2}).
    \end{aligned}
\end{equation}
This lattice admits the following symmetry:
\begin{equation}
\begin{aligned}
g:&\qty(x_1^{(A_{7})},x_2^{(A_7)},x_1^{(A_1)},x_2^{(A_1)},x_3^{(A_1)},x_4^{(A_1)};x^{(R)})\\
\mapsto&\qty(\tilde{x}_1^{(A_{7})},\tilde{x}_2^{(A_7)},-x_2^{(A_1)},-x_1^{(A_1)},-x_4^{(A_1)},-x_3^{(A_1)};x^{(R)}).
\end{aligned}
\end{equation}
The invariant lattice and its dual lattice:
\begin{equation}
    \begin{aligned}
        I=&\sqrt{2}\qty(\Lambda_R(2C_4+2A_1)+\Z(0; \sqrt{2} ,0)+\Z (0;0,\sqrt{2})+\frac{1}{2}\Z\alpha_4^{(C_4)}+\frac{1}{2}\Z\alpha_4^{(C_4)})\\
        +&\sqrt{2}\Z\qty(\frac{1}{2}(1,1)\omega_1^{(C_4)},(0,0))+\sqrt{2}\Z\qty(\frac{1}{2}(1,0)\omega_1^{(C_4)},(1,1)\omega_1^{(A_1)})\\
        +&\sqrt{2}\Z\qty((0,0),(1,0)\omega_1^{(A_1)};\frac{1}{\sqrt{2}},0)+\sqrt{2}\Z\qty(\frac{1}{2}(1,0)\omega_1^{(C_4)},(0,0);0,\frac{1}{\sqrt{2}}),\\
        I^\ast=&I+\frac{1}{\sqrt{2}}\Lambda_R(2C_4+2A_1).
    \end{aligned}
\end{equation}
By gathering the eight neutral elements and the elements $(p_L,p_R)\in I^\ast$ that satisfy $p_L^2=1, p_R^2=0$, we obtain the following representation of $2C_4+2A_1$  as massless fermions:
\begin{equation}
((\mbf{27},\mbf{1})\oplus(\mbf{1},\mbf{27}),\mbf{1},\mbf{1}), (\mbf{1},\mbf{1},(\mbf{3},\mbf{1})\oplus(\mbf{1},\mbf{3})).
\end{equation}

The gauge group is 

\begin{equation}
    \Sp(4)\times\Sp(4)\times \SU(2)\times\SU(2)\times \U(1)\times\U(1).
\end{equation}

\subsection{\texorpdfstring{$A_7+3A_3+2A_1\to C_4+C_2+A_3+A_1 ~(\#28)$}{A7+3A3+2A1 -> C4+C2+A3+A1}}

The even self-dual lattice is:
\begin{equation}
    \begin{aligned}
\Gamma_{18,2}^{(A_7+3A_3+2A_1)}=&\Lambda_R(A_7+3A_3+2A_1)+\Z(0;2,0)+\Z(0;0,2\sqrt{2})\\
+&\Z\qty(2\omega_1^{(A_7)},(2,1,1)\omega_1^{(A_3)},(0,0))+
\Z\qty(4\omega_1^{(A_7)},(2,0,0)\omega_1^{(A_3)},(1,1)\omega_1^{(A_1)})\\
+&\Z\qty(0,(1,2,0)\omega_1^{(A_3)},(0,1)\omega_1^{(A_1)};\frac{1}{2},0)
+\Z\qty(-\omega_1^{(A_{7})},(0,0,1)\omega_1^{(A_3)},(1,0)\omega_1^{(A_1)};0\frac{1}{2\sqrt{2}}).
    \end{aligned}
\end{equation}
This lattice admits the following symmetry:
\begin{equation}
\begin{aligned}
g:&\qty(x^{(A_{7})},x^{(A_3)},x_1^{(A_3)},x_2^{(A_3)},x_1^{(A_1)},x_2^{(A_1)};x^{(R)})\\
\mapsto&\qty(\tilde{x}^{(A_{7})},\tilde{x}^{(A_3)},-x_2^{(A_3)},-x_1^{(A_3)},-x_2^{(A_1)},-x_1^{(A_1)};x^{(R)}).
\end{aligned}
\end{equation}
The invariant lattice and its dual lattice  are
\begin{equation}
    \begin{aligned}
        I=&\sqrt{2}\qty(\Lambda_R(C_4+C_2+A_3+A_1)+\Z(0;\sqrt{2},0)+\Z(0;0,2)+\frac{1}{2}\Z\alpha_4^{(C_4)}+\frac{1}{2}\Z\alpha_2^{(C_2)})\\
        +&\sqrt{2}\Z\qty(\frac{1}{2}\omega_4^{(C_4)},0,2\omega_1^{(A_3)},0)+\sqrt{2}\Z\qty(\frac{1}{2}\omega_4^{(C_4)},\frac{1}{2}\omega_2^{(C_2)},0,\omega_1^{(A_1)})\\
        +&\sqrt{2}\Z\qty(0,\frac{1}{2}\omega_2^{(C_2)},0,0;\frac{1}{\sqrt{2}},0)+\sqrt{2}\Z\qty(\frac{1}{2}\omega_4^{(C_4)},\frac{1}{2}\omega_2^{(C_2)},-\omega_1^{(A_3)},0;0,\frac{1}{2}),\\
        I^\ast=&I+\frac{1}{\sqrt{2}}\Lambda_R(C_4+C_2+A_3+A_1)
    \end{aligned}
\end{equation}
By gathering the eight neutral elements and the elements $(p_L,p_R)\in I^\ast$ that satisfy $p_L^2=1, p_R^2=0$, we obtain the following representation of $C_4+C_2+A_3+A_1$  as massless fermions:
\begin{equation}
(\mbf{27},\mbf{1},\mbf{1},\mbf{1}),~(\mbf{1},\mbf{5},\mbf{1},\mbf{1}), ~(\mbf{1},\mbf{1},\mbf{15},\mbf{1}),~(\mbf{1},\mbf{1},\mbf{1},\mbf{3}). ~
\end{equation}

The gauge group is 

\begin{equation}
    \Sp(4)\times\Sp(2)\times\SU(2) \times \frac{\SU(4)\times \U(1)}{\Z_2}\times \U(1).
\end{equation}

\subsection{\texorpdfstring{$2A_9\to2C_5 ~(\#54)$}{2A9 -> 2C5}}
The even self-dual lattice is:
\begin{equation}
    \begin{aligned}
\Gamma_{18,2}^{(2A_9)}=&\Lambda_R(2A_9)+\Z\qty(2\omega_1^{(A_9)},4\omega_1^{(A_9)})+\Z(0;\sqrt{2},0)+\Z(0;0,\sqrt{2})\\
+&\Z\qty(5\omega_1^{(A_9)},0;\frac{1}{\sqrt{2}},0)+\Z\qty(0,5\omega_1^{(A_9)};0,\frac{1}{\sqrt{2}}).
    \end{aligned}
\end{equation}
This lattice admits the following symmetry:
\begin{equation}
\begin{aligned}
g:&\qty(x_1^{(A_{9})},x_2^{(A_9)};x^{(R)})\\
\mapsto&\qty(\tilde{x}_1^{(A_{9})},\tilde{x}_2^{(A_9)};x^{(R)}).
\end{aligned}
\end{equation}
The invariant lattice and its dual lattice are
\begin{equation}
    \begin{aligned}
I=&\sqrt{2}\Lambda_R(2C_5)+\Z(0;\sqrt{2},0)+\Z(0;0,\sqrt{2})\\
+&\sqrt{2}\qty(\frac{1}{2}\Z\alpha_5^{(C_5)}+\frac{1}{2}\Z\alpha_5^{(C_5)}+\Z\qty(\frac{1}{2}\omega_5^{(C_5)},0;\frac{1}{2};0)+\Z\qty(0,\frac{1}{2}\omega_5^{(C_5)};0,\frac{1}{2})),\\
I^\ast=&I+\frac{1}{\sqrt{2}}\Lambda_R(2C_5).
    \end{aligned}
\end{equation}
By gathering the eight neutral elements and the elements $(p_L,p_R)\in I^\ast$ that satisfy $p_L^2=1, p_R^2=0$, we obtain the following representation of $2C_5$  as massless fermions:
\begin{equation}
    \qty(\mbf{44},\mbf{1}),~ \qty(\mbf{1},\mbf{44}).
\end{equation}

The gauge symmetry is 

\begin{equation}
    \frac{\Sp(5)\times \U(1)}{\Z_2}\times \frac{\Sp(5)\times \U(1)}{\Z_2}.
\end{equation}

\subsection{\texorpdfstring{$A_{11}+2A_2+A_3\to C_6+A_2+C_2~(\#83)$}{A11+2A2+A3 -> C6+A2+C2}}

The even self-dual lattice is
\begin{equation}
    \begin{aligned}
\Gamma_{18,2}^{(A_{11}+2A_2+A_3)}=&\Lambda_R(A_{11}+2A_2+A_3)+\Z\qty(10\omega_1^{(A_{11})},(1,1)\omega_1^{(A_2)},2\omega_1^{(A_3)})+\Lambda\qty(\begin{matrix}
    4 & 2\\
    2 & 4\\
\end{matrix})\\
+&\Z\qty(\omega_1^{(A_{11})},(0,2)\omega_1^{(A_2)},3\omega_1^{(A_3)};e^\ast_1)+\Z\qty(0,(2,1)\omega_1^{(A_2)},2\omega_1^{(A_3)};e^\ast_2),
    \end{aligned}
\end{equation}
where $e_1^\ast,e_2^\ast$ are duals of generators of the lattice $\Lambda\qty(\begin{matrix}
    4 & 2\\
    2 & 4\\
\end{matrix})$, which is a lattice with gram matrix $\qty(\begin{matrix}
    4 & 2\\
    2 & 4\\
\end{matrix})$.

This lattice admits the following symmetry:
\begin{equation}
\begin{aligned}
    g: &\qty(x^{(A_{11})},x_1^{(A_2)},x_2^{(A_2)},x^{(A_3)};x^{(R)})\\
    \mapsto &\qty(\tilde{x}^{(A_{11})},-x_2^{(A_2)},-x_1^{(A_2)},\tilde{x}^{(A_3)};x^{(R)}).
\end{aligned}
\end{equation}
The invariant lattice and its dual lattice are
\begin{equation}
    \begin{aligned}
I=&\sqrt{2}\qty(\Lambda_R(C_6+A_2+C_2)+\frac{1}{2}\Z\alpha^{(C_6)}_6+\frac{1}{2}\Z\alpha_2^{(C_2)}+\frac{1}{2}\Z\qty(\omega_6^{(C_6)},0,\omega_2^{(C_2)}))\\
        +&\Lambda\qty(\begin{matrix}
    4 & 2\\
    2 & 4\\
\end{matrix})+\Z\qty(0,\sqrt{2}\omega_1^{(A_2)},0;2e_1^\ast)+\Z\qty(0,2\sqrt{2}\omega_1^{(A_2)},\frac{1}{\sqrt{2}}\omega_2^{(C_2)};e_2^\ast),\\
I^\ast=&I+\frac{1}{\sqrt{2}}\Lambda_R(C_6+A_2+C_2).
    \end{aligned}
\end{equation}
By gathering the eight neutral elements and the elements $(p_L,p_R)\in I^\ast$ that satisfy $p_L^2=1, p_R^2=0$, we obtain the following representation of $C_6+A_2+C_2$  as massless fermions:
\begin{equation}
(\mbf{65},\mbf{1},\mbf{1}),~(\mbf{1},\mbf{8},\mbf{1}),~(\mbf{1},\mbf{1},\mbf{5}).
\end{equation}

The gauge group is

\begin{equation}
    \Sp(6)\times\frac{\SU(3)\times \Sp(2)\times \U(1)\times \U(1)}{\Z_6}.
\end{equation}

\subsection{\texorpdfstring{$A_{11}+A_5+2A_1\to C_6+C_3+A_1(\#87)$}{A11+A5+2A1 -> C6+C3+A1}}

\begin{equation}
    \begin{aligned}
\Gamma_{18,2}^{(A_{11}+A_5+2A_1)}=&\Lambda_R(A_{11}+A_5+2A_1)+\Z\qty(10\omega^{(A_{11})}_1,2\omega^{(A_{5})}_1,(1,1)\omega^{(A_{1})}_1) +\Z(0;\sqrt{2},0)+\Z(0;0,2)\\
+&\Z\qty(8\omega^{(A_{11})}_1,\omega^{(A_{5})}_1,(1,1)\omega^{(A_{1})}_1;\frac{1}{\sqrt{2}},0)+\Z\qty(3\omega^{(A_{11})}_1,3\omega^{(A_{5})}_1,(0,1)\omega^{(A_{1})}_1;0,\frac{1}{2}).\\
    \end{aligned}
\end{equation}
This lattice admits the following symmetry:
\begin{equation}
\begin{aligned}
    g: &\qty(x^{(A_{11})},x^{(A_5)},x_1^{(A_1)},x_2^{(A_1)};x^{(R)})\\
\mapsto&\qty(\tilde{x}^{(A_{11})},\tilde{x}^{(A_5)},-x_2^{(A_1)},-x_1^{(A_1)};x^{(R)}).
\end{aligned}
\end{equation}
The invariant lattice and its dual lattice are
\begin{equation}
    \begin{aligned}
I=&\sqrt{2}\qty(\Lambda_R(C_6+C_3+A_1)+\Z(0;1,0)+\Z(0;0,\sqrt{2})+\frac{1}{2}\Z\alpha_6^{(C_6)}+\frac{1}{2}\Z\alpha_3^{(C_3)})\\
        +&\sqrt{2}\Z\qty(\frac{1}{2}\omega_6^{(C_6)},0,\omega_1^{(A_1)})+\sqrt{2}\Z\qty(0,\frac{1}{2}\omega_3^{(C_3)},\omega_1^{(A_1)};\frac{1}{2},0)+\sqrt{2}\Z\qty(\frac{1}{2}\omega_6^{(C_6)},0,0;0,\frac{1}{\sqrt{2}}),\\
        I^\ast=&I+\frac{1}{\sqrt{2}}\Lambda_R(C_6+C_3+A_1).
    \end{aligned}
\end{equation}
By gathering the eight neutral elements and the elements $(p_L,p_R)\in I^\ast$ that satisfy $p_L^2=1, p_R^2=0$, we obtain the following representation of $C_6+C_3+A_1$  as massless fermions:
\begin{equation}
(\mbf{65},\mbf{1},\mbf{1}),~(\mbf{1},\mbf{14},\mbf{1}),~(\mbf{1},\mbf{1},\mbf{3}).
\end{equation}

The gauge group is 

\begin{equation}
    \Sp(6)\times \Sp(3) \times \SU(2) \times \U(1)\times \U(1).
\end{equation}

\subsection{\texorpdfstring{$A_{15}+A_3\to C_8+C_2 ~(\#108)$}{A15+A3->C8+C2}}
The even self-dual lattice is:
\begin{equation}
    \begin{aligned}
\Gamma_{18,2}^{(A_{15}+A_3)}=&\Lambda_R(A_{15}+A_3)+\Z(0;\sqrt{2},0)+\Z(0;0,\sqrt{2})+\Z\qty(4\omega_1^{(A_{15})},2\omega_1^{(A_3)};0,0)\\
+&\Z\qty(2\omega_1^{(A_{15})},\omega_1^{(A_3)};\frac{1}{\sqrt{2}},0)+\Z\qty(2\omega_1^{(A_{15})},-\omega_1^{(A_3)};0,\frac{1}{\sqrt{2}}).
    \end{aligned}
\end{equation}
This lattice admits the following symmetry:
\begin{equation}
    \begin{aligned}
        g:&\qty(x^{(A_{15})},x^{(A_3)};x^{(R)})\\
        \mapsto &\qty(\tilde{x}^{(A_{15})}, \tilde{x}^{(A_3)};x^{(R)}).
    \end{aligned}
\end{equation}
The invariant lattice $I$ and its dual lattice  $I^\ast$ are
\begin{equation}
    \begin{aligned}
I=&\sqrt{2}\Biggl(\Lambda_R(C_8+C_2)+\Z(0;1,0)+\Z(0;0,1)+\frac{1}{2}\Z\alpha_8^{(C_8)}+\Z\frac{1}{2}\alpha_2^{(C_2)}\\
+&\Z\frac{1}{2}\omega_8^{(C_8)}+\Z\qty(0,\frac{1}{2}\omega_2^{(C_2)};\frac{1}{2},\frac{1}{2})\Biggr),\\
I^\ast=&I+\frac{1}{\sqrt{2}}\Lambda(C_8+C_2).
    \end{aligned}
\end{equation}
By gathering the eight neutral elements and the elements $(p_L,p_R)\in I^\ast$ that satisfy $p_L^2=1,p_R^2=0$, we obtain the following representation of $C_8+C_2$  as massless fermions:
\begin{equation}
\qty(\mathbf{119},\mathbf{1}),~\qty(\mathbf{1},\mathbf{5}).
\end{equation}

The gauge group is

\begin{equation}
    \Sp(8)\times\Sp(2) \times \frac{\U(1)\times \U(1)}{\Z_2}
\end{equation}

where $\Z_2$ is generated by the $\Z_2$ center symmetries  of two $\U(1)$ 's.

\subsection{\texorpdfstring{$2A_3+2D_6\to 2C_2+D_6~(\#137)$}{2A3+2D6-> 2C2+D6}}
The even self-dual lattice is:
\begin{equation}
    \begin{aligned}
\Gamma^{(2A_3+2D_6)}_{18,2}=&\Lambda_R(2A_3+2D_6)+\Z(0;2,0)+\Z(0;0,2)\\
+&\Z\qty((0,2)\omega_1^{(A_3)},(1,1)\omega_6^{(D_6)})+\Z\qty((2,0)\omega_1^{(A_3)},(1,1)\omega_5^{(D_6)})\\
+&\Z\qty((1,0)\omega_1^{(A_3)},(1,0)\omega_5^{(D_5)};\frac{1}{2},0)+\Z\qty((0,1)\omega_1^{(A_3)},(0,1)\omega_6^{(D_5)};0,\frac{1}{2}).
    \end{aligned}
\end{equation}
This lattice admits the following symmetry:
\begin{equation}
    \begin{aligned}
g:&\qty(x_1^{(A_3)},x_2^{(A_3)},x_1^{(D_6)},x_2^{(D_6)};x^{(R)})\\
\mapsto&\qty(\tilde{x}_1^{(A_3)},\tilde{x}_2^{(A_3)},-x_2^{(D_6)},-x_1^{(D_6)};x^{(R)}).\\
    \end{aligned}
\end{equation}
The invariant lattice and its dual lattice are
\begin{equation}
    \begin{aligned}
        I=&\sqrt{2}\qty(\Lambda_R(2C_2+D_6)+\Z(0;\sqrt{2},0)+\Z(0;0,\sqrt{2})+ \Z\frac{1}{2}\alpha_2^{(C_2)}+ \Z\frac{1}{2}\alpha_2^{(C_2)})\\
+&\sqrt{2}\Z\qty(\qty(0,\frac{1}{2})\omega_2^{(C_2)},\omega_6^{(D_6)})+\sqrt{2}\Z\qty(\qty(\frac{1}{2},0)\omega_2^{(C_2)},\omega_5^{(D_6)})\\
+&\sqrt{2}\Z\qty(\qty(\frac{1}{2},0)\omega_2^{(C_2)},0;\frac{1}{\sqrt{2}},0)+\sqrt{2}\Z\qty(\qty(0,\frac{1}{2})\omega_2^{(C_2)},0;0,\frac{1}{\sqrt{2}}),\\
I^\ast=&I+\frac{1}{\sqrt{2}}\Lambda_R(2C_2+D_6).
    \end{aligned}
\end{equation}
By gathering the eight neutral elements and the elements $(p_L,p_R)\in I^\ast$ that satisfy $p_L^2=1, p_R^2=0$, we obtain the following representation of $2C_2+D_6$  as massless fermions:
\begin{equation}
((\mbf{5},\mbf{1})\oplus(\mbf{1},\mbf{5}),\mbf{1}),~(\mbf{1},\mbf{1},\mbf{66}).
\end{equation}

The gauge group is 

\begin{equation}
    \Sp(2)\times \Sp(2)\times \Spin(12)\times \U(1)\times\U(1).
\end{equation}

\subsection{\texorpdfstring{$3E_6\to E_6+F_4~(\#219)$}{3E_6-> E_6+F_4}}
The even self-dual lattice is:
\begin{equation}
    \begin{aligned}
\Gamma^{(3E_6)}_{18,2}=&\Lambda_R(3E_6)+\Z\qty(1,1,1)\omega_1^{(E_6)}+\Lambda\qty(\begin{matrix}
    2 & 1\\
    1 & 2\\
\end{matrix})\\
+&\Z\qty((1,-1,0)\omega_1^{(E_6)};e_1^\ast)+\Z\qty((-1,1,0)\omega_1^{(E_6)};e_2^\ast),
    \end{aligned}
\end{equation}
where $e_1,e_2$ are the generators of  $\Lambda\qty(\begin{matrix}
    2 & 1\\
    1 & 2\\
\end{matrix})$, which is a lattice with gram matrix $\qty(\begin{matrix}
    2 & 1\\
    1 & 2\\
\end{matrix})$.

This lattice admits the following symmetry:
\begin{equation}
\begin{aligned} g:&\qty(x_1^{(E_6)},x_2^{(E_6)},x^{(E_6)};x^{(R)})\
\mapsto&\qty(-x_2^{(E_6)},-x_1^{(E_6)},\tilde{x}^{(E_6)};x^{(R)}),
\end{aligned}
\end{equation}

where $\tilde{x}^{(E_6)}$ is obtained by exchanging $\alpha_1^{(E_6)}\leftrightarrow\alpha_6^{(E_6)}$ and $\alpha_3^{(E_6)}\leftrightarrow\alpha_5^{(E_6)}$ in $x^{(E_6)}$.

The invariant lattice and its dual lattice are
\begin{equation}
    \begin{aligned}
I=&\sqrt{2}\Lambda_R(E_6+F_4)+\Lambda\qty(\begin{matrix}
    2 & 1\\
    1 & 2\\
\end{matrix})+\Z\qty(\sqrt{2}\omega_1^{(E_6)},e_1^\ast)+\Z\qty(-\sqrt{2}\omega_1^{(E_6)},e_2^\ast),\\
  I^\ast=&I+\frac{1}{\sqrt{2}}\qty(\Lambda_R(E_6)+\Z\alpha_1^{(F_4)}+\Z\alpha_2^{(F_4)}).\\
    \end{aligned}
\end{equation}
By gathering the eight neutral elements and the elements $(p_L,p_R)\in I^\ast$ that satisfy $p_L^2=1, p_R^2=0$, we obtain the following representation of $E_6+F_4$  as massless fermions:
\begin{equation}
(\mbf{78},\mbf{1}),\qty(\mbf{1},\mbf{26}).
\end{equation}

The gauge group is 

\begin{equation}
    E_6\times F_4\times \U(1)\times \U(1).
\end{equation}

\subsection{\texorpdfstring{$D_4+2E_7\to B_3+E_7~(\#279)$}{D4+2E7 -> B3+E7}}
The even self-dual lattice is:
\begin{equation}
    \begin{aligned}
\Gamma^{(D_4+2E_7)}_{18,2}=&\Lambda_R(D_4+2E_7)+\Z\qty(\omega_4^{(D_4)},\omega_7^{(E_7)},\omega_7^{(E_7)})+\Z(0;\sqrt{2},0)+\Z(0;0,\sqrt{2})\\
&+\Z\qty(\omega_3^{(D_4)},\omega_7^{(E_7)},0;\frac{1}{\sqrt{2}},0)+\Z\qty(\omega_3^{(D_4)},0,\omega_7^{(E_7)};0,\frac{1}{\sqrt{2}}).
    \end{aligned}
\end{equation}
This lattice admits the following symmetry:
\begin{equation}
    \begin{aligned}
g:&\qty(x^{(D_{4})},x_1^{(E_7)},x_2^{(E_7)};x^{(R)}),\\
\mapsto&\qty(\tilde{x}^{(D_{4})},-x_2^{(E_7)},-x_1^{(E_7)};x^{(R)}),
    \end{aligned}
\end{equation}
where $\tilde{x}^{(D_4)}$ is obtained by exchanging $\alpha_1^{(D_4)}\leftrightarrow\alpha_3^{(D_4)}$ in $x^{(D_4)}$.

The invariant lattice and its dual lattice are
\begin{equation}
    \begin{aligned}
I=&\Z\alpha_1^{(B_3)}+\Z\alpha_2^{(B_3)}+\Z 2\alpha_3^{(B_3)}+\sqrt{2}\Lambda_R(E_7)+\Z(0;\sqrt{2},0)+\Z(0;0,\sqrt{2})\\
+&\Z\qty(\alpha_3^{(B_3)},\sqrt{2}\omega_7^{(E_7)})+\Z\qty(0,\sqrt{2}\omega_7^{(E_7)};\frac{1}{\sqrt{2}},-\frac{1}{\sqrt{2}}),\\
I^\ast=&I+\frac{1}{\sqrt{2}}\Lambda_R(E_7)+\Z\qty(\frac{1}{2}\qty(\alpha_1^{(B_3)}+\alpha_3^{(B_3)})+\frac{1}{\sqrt{2}}\omega_7^{(E_7)};\frac{1}{\sqrt{2}},0).
    \end{aligned}
\end{equation}
By gathering the eight neutral elements and the elements $(p_L,p_R)\in I^\ast$ that satisfy $p_L^2=1, p_R^2=0$, we obtain the following representation of $B_3+E_7$  as massless fermions:
\begin{equation}
(\mbf{7},\mbf{1}),\qty(\mbf{1},\mbf{133}).
\end{equation}

The gauge group is 

\begin{equation}
    \frac{E_7\times \Spin(6)\times \U(1)}{\Z_2} \times \U(1)
\end{equation}

\subsection{\texorpdfstring{$2E_8+A_2\to E_8+A_2~(\#297)$}{2E8+A2->E8+A2}}

The even self-dual lattice is:
\begin{equation}
    \begin{aligned}
\Gamma^{(2E_8+A_2)}_{18,2}=&\Lambda_R(2E_8+A_2)+\Lambda\qty(\begin{matrix}
    2 & 1\\
    1 & 2\\
\end{matrix})+\Z\qty(\omega_1^{(A_2)},e_1^\ast)+\Z\qty(-\omega_1^{(A_2)},e_2^\ast),
    \end{aligned}
\end{equation}
where $e_1,e_2$ are generators of  $\Lambda\qty(\begin{matrix}
    2 & 1\\
    1 & 2\\
\end{matrix})$, which is a lattice with gram matrix $\qty(\begin{matrix}
    2 & 1\\
    1 & 2\\
\end{matrix})$.
This lattice admits the following symmetry:
\begin{equation}
    \begin{aligned}
g:&\qty(x_1^{(E_8)},x_2^{(E_8)},x^{(A_2)};x^{(R)})\\
        \mapsto&\qty(-x_2^{(E_8)},-x_1^{(E_8)},x^{(A_2)};x^{(R)}).
    \end{aligned}
\end{equation}
The invariant lattice and its dual lattice are
\begin{equation}
\begin{aligned}
    I=&\sqrt{2}\Lambda_R(E_8)+\Lambda\qty(\begin{matrix}
    2 & 1\\
    1 & 2\\
\end{matrix})+\Lambda_R(A_2)+\Z\qty(\omega_1^{(A_2)},e_1^\ast)+\Z\qty(-\omega_1^{(A_2)},e_2^\ast),\\
I^\ast=&I+\frac{1}{\sqrt{2}}\Lambda_R(E_8).
\end{aligned}
\end{equation}
By gathering the eight neutral elements and the elements $(p_L,p_R)\in I^\ast$ that satisfy $p_L^2=1, p_R^2=0$, we obtain the following representation of $E_8+A_2$  as massless fermions:
\begin{equation}
(\mbf{248},\mbf{1}).
\end{equation}

The gauge group is 

\begin{equation}
    E_8\times \frac{\SU(3)\times \U(1)\times \U(1)}{\Z_3\times\Z_3}.
\end{equation}

\subsection{Non-supersymmetric branes}
In this section, we have seen the various points of gauge enhancement with the $\mathbb{Z}_2$ outer automorphism symmetry $g$, as summarized in Table~\ref{tab:8d_summary}.
This indicates that the disconnected gauge groups of the supersymmetric heterotic strings.
Applying the no global symmetry/cobordism conjecture, the existence of the codimension-two non-supersymmetric brnae is predicted in supersymmetric heterotic theory.
It is known that, in 10d heterotic theory, such a brane is characterized by the monodromy exchanging two $E_8$s~\cite{Kaidi:2023tqo}.
Similarly, in \cite{Hamada:2025JHEP}, we find that the monodromy involves the exchange as well as the charge conjugation symmetry of $A_n$.
In this paper, we have found the new features that the monodromy corresponding to the folding of $D_4$ and $E_6$ Dynkin diagrams.

\section{\texorpdfstring{$E_8$}{E8} string theory on \texorpdfstring{$T^d$}{T^d}}\label{sec:E8}
In this section, we consider $S^1$ and $T^2$ compactifications of the ten-dimensional non-supersymmetric $E_8$ string theory, and investigate the moduli space of maximal gauge enhancement.
We confirm that the gauge enhancement is precisely the same as the result in section~\ref{sec:8d_folding}. Note that the $E_8$ string theory on $T^d$ always has a moduli-independent tachyon that cannot be removed by turning on moduli \cite{DeFreitas:2024ztt}.

\subsection{Construction}\
The ten-dimensional non-supersymmetric $E_8$ string theory can be constructed by orbifolding the ten dimensional supersymmetric $E_{8}\times E'_{8}$ string theory \eqref{eq:SUSY_partition_function} by $g=R(-1)^{F}$, where $R$ is the outer automorphism of the $E_8\times E'_8$ lattice $\Gamma_{E_{8}}\oplus\Gamma_{E'_{8}}$ and $F$ is the spacetime fermion number. Its partitionfunction can be expressed as the sum of these blockes (see e.g. \cite{DeFreitas:2024ztt} and Appendix B of \cite{Nakajima:2023zsh}):
\begin{equation}
	\begin{aligned}
		Z^{(10)}(1,1)&=\frac{1}{2}Z_{B}^{(8)}(\bar{V}_{8}-\bar{S}_{8})\times \frac{1}{\eta^{16}}\sum_{\pi\in\Gamma_{E_{8}}}\sum_{\pi'\in\Gamma_{E'_{8}}}q^{\frac{1}{2}(\pi^{2}+\pi'^{2})},\\
		Z^{(10)}(1,g)&=\frac{1}{2}Z_{B}^{(8)}(\bar{V}_{8}+\bar{S}_{8})\times \frac{1}{\eta^{16}}\left(\frac{2\eta^3}{\theta_{2}}\right)^{4}\sum_{\pi\in\sqrt{2}\Gamma_{E_{8}}}q^{\frac{1}{2}\pi^{2}},\\
		Z^{(10)}(g,1)&=\frac{1}{2}Z_{B}^{(8)}(\bar{O}_{8}-\bar{C}_{8})\times \frac{1}{\eta^{16}}\left(\frac{\eta^3}{\theta_{4}}\right)^{4}\sum_{\pi\in\frac{1}{\sqrt{2}}\Gamma_{E_{8}}}q^{\frac{1}{2}\pi^{2}},\\
		Z^{(10)}(g,g)&=\frac{1}{2}Z_{B}^{(8)}(\bar{O}_{8}+\bar{C}_{8})\times \frac{1}{\eta^{16}}\left(\frac{\eta^3}{\theta_{3}}\right)^{4}\sum_{\pi\in\frac{1}{\sqrt{2}}\Gamma_{E_{8}}}q^{\frac{1}{2}\pi^{2}}e^{\pi i \pi^2}.
	\end{aligned}
\end{equation}
Compactifying this theory on $T^{d}$, we obtain
\begin{equation}
	\begin{aligned}\label{eq:E8_partition}
		Z^{(10-d)}(1,1)&=\frac{1}{2}Z_{B}^{(8-d)}(\bar{V}_{8}-\bar{S}_{8})\times \frac{1}{\eta^{16+d}\bar{\eta}^d}\sum_{(p_{-},P)\in\Gamma_{16+d,d}}q^{\frac{1}{2}(p_{-}^{2}+P_{L}^{2})}\bar{q}^{\frac{1}{2}p_{R}^{2}},\\
		Z^{(10-d)}(1,g)&=\frac{1}{2}Z_{B}^{(8-d)}(\bar{V}_{8}+\bar{S}_{8})\times \frac{1}{\eta^{16+d}\bar{\eta}^d}\left(\frac{2\eta^3}{\theta_{2}}\right)^{4}\sum_{P\in I_{8+d,d}}q^{\frac{1}{2}P_{L}^{2}}\bar{q}^{\frac{1}{2}p_{R}^{2}},\\
		Z^{(10-d)}(g,1)&=\frac{1}{2}Z_{B}^{(8-d)}(\bar{O}_{8}-\bar{C}_{8})\times \frac{1}{\eta^{16+d}\bar{\eta}^d}\left(\frac{\eta^3}{\theta_{4}}\right)^{4}\sum_{P\in I^{*}_{8+d,d}}q^{\frac{1}{2}P_{L}^{2}}\bar{q}^{\frac{1}{2}p_{R}^{2}},\\
		Z^{(10-d)}(g,g)&=\frac{1}{2}Z_{B}^{(8-d)}(\bar{O}_{8}+\bar{C}_{8})\times \frac{1}{\eta^{16+d}\bar{\eta}^d}\left(\frac{\eta^3}{\theta_{3}}\right)^{4}\sum_{P\in I^{*}_{8+d,d}}q^{\frac{1}{2}P_{L}^{2}}\bar{q}^{\frac{1}{2}p_{R}^{2}}e^{\pi i P^2}.
	\end{aligned}
\end{equation}
The elements of $\Gamma_{16+d,d}$ are expressed in terms of moduli such as 
 the metric $G=ee^t$,\footnote{Here $e$ is the vierbein, and $t$ is the transpose.} the anti-symmetric tensor $B$, the Wilson lines $A=(a,a)$\footnote{By the consistency with orbifolding, Wilson lines are restricted to take the form as $A=(a,a)$.}:
\begin{equation}\label{momenta}
    \begin{aligned}
        p_{-}&=\frac{1}{\sqrt{2}}\pi_{-},\\
        p_{+}&=\frac{1}{\sqrt{2}}\left(\pi_{+}+2wa\right),\\
        p_{L}&=\frac{1}{\sqrt{2}}\left(n+w(2G-E)-\pi_{+}\cdot a\right)~e^{-t}=p_{R}+\sqrt{2}we,\\
        p_{R}&=\frac{1}{\sqrt{2}}\left(n-wE-\pi_{+}\cdot a\right)~e^{-t},
    \end{aligned}
\end{equation}
where we define $E=G+B+a\cdot a$. We have adopted the notations of row vectors 
\begin{equation}
   \pi_{\pm}:=\pi\pm\pi'\in\Gamma_{E_{8}},
   ~~w\in\mathbb{Z}^{d},~~n\in\mathbb{Z}^{d}.
\end{equation}
The elements of $I_{8+d,d}$ and $I^{*}_{8+d,d}$ are written by $P=(P_{L};p_{R})=(p_{+},p_{L};p_{R})$ which satisfies
\begin{equation}
    P^2=P_{L}^2-p_{R}^2=\frac{1}{2}\pi_{+}^{2}+2wn^{t}.
\end{equation}
Note that $\pi_{+}\in2\Gamma_{E_{8}}$ for $P\in I_{8+d,d}$.


\subsection{Massless States and Maximal Enhancements of 9d theories}\label{sec:9d_maximal}
\subsubsection{Massless Vectors}
To identify the maximal enhancements, we first focus on the massless gauge bosons which come from the untwisted sectors combined with the $D_4$ character vector conjugacy class in 9d theories. The massless conditions for the gauge bosons which depend on the moduli are the same as those for the untwisted sectors of CHL string (for details, see e.g. \cite{Font:2021uyw}):
 \begin{subequations}\label{massless condition}
 	\begin{align}
 		P_{L}^{2}&=\frac{1}{2}\pi_{+}^{2}+2wn=
        \begin{cases}
            1 & \text{with}~~ \pi_{-}^2=1, \\
            2 & \text{with}~~ \pi_{-}=0,
        \end{cases}\\
 		p_{R}&=n -wE -\pi_{+}\cdot a=0,
 	\end{align}
 \end{subequations}
where $E=R^2+a^2$. The solutions for $P_{L}^{2}=1$ correspond to short roots while those for $P_{L}^{2}=2$ correspond to long roots. Note that for the $P_{L}^{2}=2$ case, $\pi_{+}$ is more restricted to $\pi_{+}\in2\Gamma_8$. In the 9d CHL string, $n$ must be even and there are no solutions for $P_{L}^{2}=2$ in the untwisted sectors with $d=1$, which does not allow non-simply raced gauge symmetries in the 9d CHL string. The $E_8$ string on $S_1$, however, has the solutions which give non-simply laced gauge enhancements even in 9d theories with specific moduli.

Let us consider the simplest case where $a=0$. 
The condition $p_R=0$ gives $n=wR^2$, thus we get the solutions for any $R$:
\begin{align}\label{non-zero root solutions}
	\pi_{+}^2=2,~~w=n=0,
\end{align}
which satisfy $P_{L}^{2}=1$ and correspond to the non-zero roots of $E_8$. One can also find the solutions for $P_{L}^{2}=2$ when $R=1$, which give long roots:
\begin{align}\label{C1 condition}
	\pi_{+}=0,~~w=n=\pm 1.
\end{align}
Note that these states cannot be found in the 9d CHL string. From these solutions, the gauge group is $E_{8}\times C_{1}$, rather than $E_{8}\times A_{1}$, because of existing the long roots in the case with $R=1, a=0$.

\begin{table}[h]
    \centering
    \begin{tabular}{|c|c||c|c|} \hline
        $~i~$ & $\kappa_i$ & $\alpha_i$ & $\omega_i$ \\ \hline
        1 & $3$ & $(1,\minus1,0,0,0,0,0,0)$ & $\minus(\minus\frac{1}{2},\frac{1}{2},\frac{1}{2},\frac{1}{2},\frac{1}{2},\frac{1}{2},\frac{1}{2},\minus\frac{5}{2})$ \\
        2 & $6$ & $(0,1,\minus1,0,0,0,0,0)$ & $\minus(0,0,1,1,1,1,1,\minus5)$ \\
        3 & $5$ & $(0,0,1,\minus1,0,0,0,0)$ & $\minus(0,0,0,1,1,1,1,\minus4)$ \\
        4 & $4$ & $(0,0,0,1,\minus1,0,0,0)$ & $\minus(0,0,0,0,1,1,1,\minus3)$ \\
        5 & $3$ & $(0,0,0,0,1,\minus1,0,0)$ & $\minus(0,0,0,0,0,1,1,\minus2)$ \\
        6 & $2$ & $(0,0,0,0,0,1,\minus1,0)$ & $(0,0,0,0,0,0,\minus1,1)$ \\
        7 & $4$ & $\minus(1,1,0,0,0,0,0,0)$ & $\minus(\frac{1}{2},\frac{1}{2},\frac{1}{2},\frac{1}{2},\frac{1}{2},\frac{1}{2},\frac{1}{2},\minus\frac{7}{2})$ \\
        8 & $2$ & $(\frac{1}{2},\frac{1}{2},\frac{1}{2},\frac{1}{2},\frac{1}{2},\frac{1}{2},\frac{1}{2},\frac{1}{2})$ & $(0,0,0,0,0,0,0,2)$ \\
        0 & $1$ & $(0,0,0,0,0,0,1,\minus1)$ & $(0,0,0,0,0,0,0,0)$ \\ \hline
    \end{tabular}
    \caption{Kac marks, Simple roots and fundamental weights of $E_8$.}
    \label{table_E8_simple_roots}
\end{table}

Specifying the massless vectors in the case of nonzero $a$ is also straightforward. Here we briefly comment on the outline. For short loots, the same solutions as in the 9d CHL can be found if the Wilson line takes the form as $a=\omega_{i}/\kappa_{i}$ where $\omega_{i}$ and $\kappa_{i}$ are the fundamental weight and the Kac mark of $E_{8}$ shown in Table \ref{table_E8_simple_roots}. 

For example, we find
\begin{align}
    \pi_{+}^2=2,~~w=n=0,~~\pi_{+}\cdot \omega_{i}=0.
\end{align}
From the first and third conditions, $\pi_{+}=\alpha_{j},~j\neq i$ can survive, but $\pi_{+}=\alpha_{i}$ is projected out, where $\alpha_{i}$ is the simple roots of $E_{8}$. For long roots, we always obtain (\ref{C1 condition}) when $E=1$. 

For solving \eqref{massless condition} with the moduli $E=1, a=\omega_{i}/\kappa_{i}$, we finally obtain the eight maximal enhanced gauge groups shown in Table \ref{table_maximal_enhancement_S1}.
Note that we cannot obtain a maximal enhancement with $a=\omega_{8}/2$ which is not consistent with $E=1$, since $a^2=1$ gives $R=0$. This leads to the technique of the generalized Dynkin diagram for the 9d theories that can identify the maximal enhancements, discussed in subsection \ref{sec:9d_Dynkin}.

\begin{table}[h]
    \centering
    \begin{tabular}{|c||c|c||c|c|} \hline
        $i$ & $E_8\times E'_8$ & $E_8$ & $E$ & $a$ \\ \hline \hline
        $1$ & $A_{17}$ & $C_9$ & $1$ & $\frac{\omega_1}{3}$  \\ \hline
        $2$ & $2A_1 + 2A_2 + A_{11}$ & $C_6 + A_2 + A_1$ & $1$ & $\frac{\omega_2}{6}$  \\ \hline
        $3$ & $2A_4 + A_9$ & $C_5 + A_4 $ & $1$ & $\frac{\omega_3}{5}$  \\ \hline
        $4$ & $A_7 + 2D_5$ & $C_4 + D_5$ & $1$ & $\frac{\omega_4}{4}$  \\ \hline
        $5$ & $A_5 + 2E_6$ & $C_3 + E_6$ & $1$ & $\frac{\omega_5}{3}$  \\ \hline
        $6$ & $A_3 + 2E_7$ & $C_2 + E_7$ & $1$ & $\frac{\omega_6}{2}$  \\ \hline
        $7$ & $2A_1 + A_{15}$ & $C_8 + A_1$ & $1$ & $\frac{\omega_7}{4}$  \\ \hline
        $0$ & $A_1 + 2E_8$ & $C_1 + E_8$ & $1$ & $\omega_0$  \\ \hline
    \end{tabular}
    \caption{Maximal enhancements and corresponding moduli in the the $E_8\times E'_8$ and $E_8$ string on $S^1$, where Wilson line $a$ is expressed by data in Table \ref{table_E8_simple_roots}. The latter ones can be obtained by deleting the $i$-th node of the 9d EDD shown in Figure \ref{Fig_9d_EDD}.}
    \label{table_maximal_enhancement_S1}
\end{table}

\subsubsection{Massless Matters}
Next we consider the matter spectra. In particular, to compare the results in section \ref{sec:9d_review}, we focus on the massless co-spinors which come from the untwisted sectors combined with the $D_4$ character co-spinor conjugacy class in 9d theories. From \eqref{eq:E8_partition}, co-spinor states are paired with the dual lattice. The massless conditions for the co-spinors are then given by
\begin{subequations}\label{eq:massless_condition_matter}
 	\begin{align}
 		P_{L}^{2}&=\frac{1}{2}\pi_{+}^{2}+2wn=1,\\
 		p_{R}&=n -wE -\pi_{+}\cdot a=0.
 	\end{align}
 \end{subequations}
These conditions are nothing but \eqref{massless condition} with $P_{L}^2=1$, which correspond to the short roots. Then the massless matters are expressed by the quasi-minuscule representation of $C_n$. We finally obtain the same results as in Table \ref{tab:9d_summary} by solving \eqref{eq:massless_condition_matter} with the moduli $E=1, a=\omega_{i}/\kappa_{i},~i\neq8$.

\subsection{Generalized Dynkin Diagram for \texorpdfstring{$S^1$}{S1}}\label{sec:9d_Dynkin}
We propose the generalized Dynkin diagram (EDD) for the $E_8$ string on $S_1$, which can be used for the gauge group with the maximal rank. The difference from the EDD for the 9d CHL string \cite{Font:2021uyw} is the $\texttt{C}$ node that gives the condition of $E$.

We define the charge vector
\begin{align}
    Z=\ket{2w,n;\pi_{+}},
\end{align}
and the inner product of charge vectors is defined through the metric
\begin{equation}
    \eta=
    \begin{pmatrix}
        0 & 1 & 0 \\
        1 & 0 & 0 \\
        0 & 0 & \boldsymbol{1}_{8}
    \end{pmatrix}.
\end{equation}
Note that $Z^2=\pi_{+}^2+4wn=2P^2$.
The nodes 1 through 8 are the simple roots of $E_{8}$
\begin{align}
    Z_{i}=\ket{0,0;\alpha_{i}},~~~i=1,\ldots,8,
\end{align}
and the 0 node is the lowest root of $E_{8}$ with non-zero momentum
\begin{align}
    Z_{0}=\ket{0,-1;\alpha_{0}}.
\end{align}
The lowest root $\alpha_{0}$ is expressed as
\begin{align}
    \alpha_{0}=-\sum_{i=1}^{8}\kappa_{i}\alpha_{i}.
\end{align}
Finally, the $\texttt{C}$ node is
\begin{align}
    Z_{\texttt{C}}=\ket{2,1;0^8},
\end{align}
which corresponds to the long root and differs the $\texttt{C}$ node of the 9d CHL string $\ket{1,1;0}$. The difference in the first component comes from the fact that the winding number can take a half-integer value in the 9d CHL string while that is impossible in the $E_8$ string on $S_1$. That also leads to the condition $E=1$ for the maximal enhancements, unlike $E=2$ in the 9d CHL string.

\begin{figure}[h]
    \begin{center}
        \begin{tikzpicture}[scale=1.0]
            \tikzset{double circle/.style={draw, circle, double, thick, minimum size=10pt, inner sep=1pt}}
            
    \node[thick, circle, draw, fill=white] (1) at (0,0) {};
    \node[thick, circle, draw, fill=white] (2) at (1,0) {};
    \node[thick, circle, draw, fill=white] (3) at (2,0) {};
    \node[thick, circle, draw, fill=white] (4) at (3,0) {};
    \node[thick, circle, draw, fill=white] (5) at (4,0) {};
    \node[thick, circle, draw, fill=white] (6) at (5,0) {};
    \node[thick, circle, draw, fill=white] (7) at (1,1) {};
    \node[thick, circle, draw, fill=white] (8) at (1,2) {};
    \node[thick, circle, draw, fill=yellow] (0) at (6,0) {};
    \node[double circle, fill=yellow] (C) at (7,0) {};

    \draw[thick] (1) -- (2);
    \draw[thick] (2) -- (3);
    \draw[thick] (3) -- (4);
    \draw[thick] (4) -- (5);
    \draw[thick] (5) -- (6);
    \draw[thick] (2) -- (7);
    \draw[thick] (7) -- (8);
    \draw[thick] (6) -- (0);
    \draw[double distance=2pt, thick] (0) -- (C);
    \draw[thick] (6.6,0.15) -- (6.4,0.0) -- (6.6,-0.15);

    \node[below=0.3cm] at (1) {1};
    \node[below=0.3cm] at (2) {2};
    \node[below=0.3cm] at (3) {3};
    \node[below=0.3cm] at (4) {4};
    \node[below=0.3cm] at (5) {5};
    \node[below=0.3cm] at (6) {6};
    \node[left=0.3cm] at (7) {7};
    \node[left=0.3cm] at (8) {8};
    \node[below=0.3cm] at (0) {0};
    \node[below=0.3cm] at (C) {$\texttt{C}$};
        \end{tikzpicture}
        \caption{Generalized Dynkin diagram for 9d theories.}\label{Fig_9d_EDD}
    \end{center}
\end{figure}
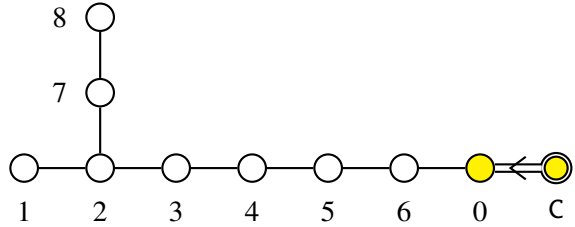

We show the generalized Dynkin diagram for the $E_8$ string theory on $S^1$ in Figure \ref{Fig_9d_EDD}. We colored the nodes that have momentum and/or winding charge and the double border of the node implies that it corresponds to a vector with $P^2=2$. From this diagram, we can determine the maximal enhancements of this theory by deleting one node and we can also read off the corresponding moduli as shown in Table \ref{table_maximal_enhancement_S1}. Note that the case of deleting the node $8$ is excluded since it gives the unphysical condition such that $a^2=\left(\frac{\omega_8}{2}\right)^2=1$ and $R^2=E-a^2=0$. These results are completely the same as in \cite{Hamada:2025JHEP}.

\begin{figure}[h]
    \begin{center}
        \begin{tikzpicture}[scale=0.8]
    \node[thick, circle, draw, fill=white] (1) at (0,0) {};
    \node[thick, circle, draw, fill=white] (2) at (1,0) {};
    \node[thick, circle, draw, fill=white] (3) at (2,0) {};
    \node[thick, circle, draw, fill=white] (4) at (3,0) {};
    \node[thick, circle, draw, fill=white] (5) at (4,0) {};
    \node[thick, circle, draw, fill=white] (6) at (5,0) {};
    \node[thick, circle, draw, fill=white] (7) at (1,1) {};
    \node[thick, circle, draw, fill=white] (8) at (1,2) {};
    \node[thick, circle, draw, fill=yellow] (0) at (6,0) {};
    \node[thick, circle, draw, fill=yellow] (C) at (7,0) {};
    \node[thick, circle, draw, fill=white] (1') at (14,0) {};
    \node[thick, circle, draw, fill=white] (2') at (13,0) {};
    \node[thick, circle, draw, fill=white] (3') at (12,0) {};
    \node[thick, circle, draw, fill=white] (4') at (11,0) {};
    \node[thick, circle, draw, fill=white] (5') at (10,0) {};
    \node[thick, circle, draw, fill=white] (6') at (9,0) {};
    \node[thick, circle, draw, fill=white] (7') at (13,1) {};
    \node[thick, circle, draw, fill=white] (8') at (13,2) {};
    \node[thick, circle, draw, fill=yellow] (0') at (8,0) {};
    
    \draw[thick] (1) -- (2);
    \draw[thick] (2) -- (3);
    \draw[thick] (3) -- (4);
    \draw[thick] (4) -- (5);
    \draw[thick] (5) -- (6);
    \draw[thick] (2) -- (7);
    \draw[thick] (7) -- (8);
    \draw[thick] (6) -- (0);
    \draw[thick] (0) -- (C);
    \draw[thick] (1') -- (2');
    \draw[thick] (2') -- (3');
    \draw[thick] (3') -- (4');
    \draw[thick] (4') -- (5');
    \draw[thick] (5') -- (6');
    \draw[thick] (2') -- (7');
    \draw[thick] (7') -- (8');
    \draw[thick] (6') -- (0');
    \draw[thick] (0') -- (C);
    \draw[dotted, thick, blue] (7, -1.3) -- (7, 2.7);
    \draw[thick, blue, <->] (5,1.5) to [out=45,in=135] (9,1.5);

    \node[below=0.3cm] at (1) {1};
    \node[below=0.3cm] at (2) {2};
    \node[below=0.3cm] at (3) {3};
    \node[below=0.3cm] at (4) {4};
    \node[below=0.3cm] at (5) {5};
    \node[below=0.3cm] at (6) {6};
    \node[left=0.3cm] at (7) {7};
    \node[left=0.3cm] at (8) {8};
    \node[below=0.3cm] at (0) {0};
    \node[below=0.3cm] at (C) {\texttt{C}};
    \node[below=0.3cm] at (1') {1'};
    \node[below=0.3cm] at (2') {2'};
    \node[below=0.3cm] at (3') {3'};
    \node[below=0.3cm] at (4') {4'};
    \node[below=0.3cm] at (5') {5'};
    \node[below=0.3cm] at (6') {6'};
    \node[left=0.3cm] at (7') {7'};
    \node[left=0.3cm] at (8') {8'};
    \node[below=0.3cm] at (0') {0'};
        \end{tikzpicture}
        \caption{EDD for the 9d theories can be obtained by folding the SUSY one.}\label{Fig_folding_9d_EDD}
    \end{center}
\end{figure}
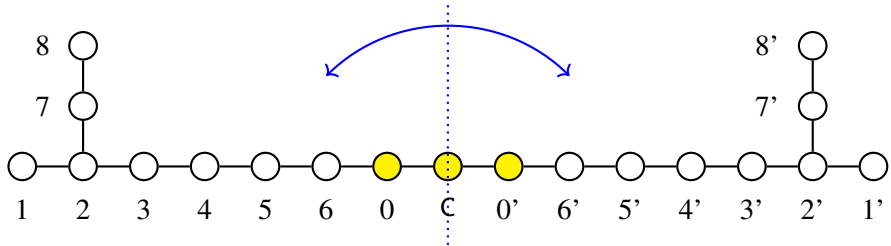

This diagram can also be obtained by folding the generalized Dynkin diagram for the heterotic superstrings on $S^1$ \cite{Font:2020rsk}, as shown in Figure \ref{Fig_folding_9d_EDD}. That is consistent with the statement that this theory can be constructed from the 9d heterotic $E_8\times E_8$ superstring by orbifolding of the outer automorphism as well as the fermion parity. Note that the nodes $8$ and $8'$ cannot be removed at the same time in the SUSY EDD since it does not give a possible Dynkin diagram. That is also consistent with the fact that node 8 cannot be removed in the EDD in Figure \ref{Fig_9d_EDD}. From this operation, we can see that removing the $i$-th node in Figure \ref{Fig_9d_EDD} corresponds to removing the $i$- and $i'$-th node in the SUSY Dynkin diagram, which implies that the maximal enhancements of $E_8$ string can be obtained by folding the only eight maximal enhancements of SUSY theoris in Table 11 of \cite{Font:2020rsk}, where the Wilson line takes the form $A=(a,a)$.

\subsection{Massless States and Maximal Enhancements of 8d theories}\label{sec:8d_maximal}
\subsubsection{Massless Vectors}
Let us consider the massless vectors in 8d theories. The massless conditions for 8d are the generalized ones of \eqref{massless condition}:
\begin{subequations}\label{massless condition2}
 	\begin{align}
 		P_{L}^{2}&=\frac{1}{2}\pi_{+}^{2}+2w^1n_1+2w^2n_2=
        \begin{cases}
            1 & \text{with}~~ \pi_{-}^2=1, \\
            2 & \text{with}~~ \pi_{-}=0,
        \end{cases}\\
 		p_{R,i}&=n_i -w^j E_{ij} -\pi_{+}\cdot a_{i}=0,~~~i=1,2.
 	\end{align}
 \end{subequations}
In order to obtain the 8d maximal enhancements, we use the data for the gauge groups with maximal rank of the 8d SUSY heterotic string theory and the corresponding moduli, as shown in Table 12 of \cite{Font:2020rsk}. The criteria for choosing the theories is that both two Wilson lines of SUSY theories are expressed as $A_{1}=(a_1,a_1),~A_{2}=(a_2,a_2)$, and use the same moduli $a_1, a_2$ and $E$. Solving \eqref{massless condition2} with these moduli, we can then obtain the 22 maximal enhanced theories, as shown in Table \ref{table_maximal_enhancement_T2}. The maximally enhanced gauge groups in Table \ref{table_maximal_enhancement_T2} are consistent with the results obtained by folding the lattice in section \ref{sec:8d_folding}. In the 8d SUSY theories, $\#1$ and $\#2$ were derived using the Fixed Wilson line algorithm, while the others were obtained using the method of generalized Dynkin diagrams. In subsection \ref{sec:8d_Dynkin}, we show that those 20 theories can be obtained with the method of EDDs.

\begin{table}[t]
    \centering
    \begin{tabular}{|c||c|c||cccc|c|c|} \hline
        $\#$ & $E_8\times E'_8$ & $E_8$ & $E_{11}$ & $E_{12}$ & $E_{21}$ & $E_{22}$ & $a_1$ & $a_2$ \\ \hline \hline
        $1$ & $6A_3$ & $2C_2+2A_3$ & $1$ & $\frac{1}{2}$ & $0$ & $1$ & $\frac{\omega_6}{2}$ & $\frac{\omega_2}{4}-\frac{3\omega_6}{4}$ \\ \hline
        $2$ & $2A_1+4A_4$ & $2C_1+2A_4$ & $1$ & $-\frac{1}{5}$ & $\frac{1}{5}$ & $1$ & $\frac{\omega_3}{5}$ & $w_1 -\frac{3\omega_3}{5}$ \\ \hline
        $5$ & $4A_2 + 2A_5$ & $2A_2+2C_3$ & $1$ & $0$ & $0$ & $1$ & $\frac{\omega_5}{3}$ & $\frac{\omega_2}{3}-\frac{2\omega_5}{3}$\\ \hline
        $6$ & $A_3 + 3A_5$ & $C_2 + C_3 + A_5$ & $1$ & $0$ & $0$ & $1$ & $\frac{\omega_5}{3}$ & $\frac{\omega_1}{2}-\frac{\omega_5}{2}$ \\ \hline
        $25$ & $4A_1 + 2A_7$ & $2A_1 + 2C_4$ & $1$ & $0$ & $0$ & $1$ & $\frac{\omega_4}{4}$ & $\frac{\omega_2}{2}-\frac{3\omega_4}{4}$\\ \hline
        $28$ & $2A_1 + 3A_3 + A_7$ & $A_1 + C_2 + A_3 + C_4$ & $1$ & $0$ & $0$ & $1$ & $\frac{\omega_4}{4}$ & $\frac{\omega_4}{2}-\frac{\omega_7}{2}$\\ \hline
        $54$ & $2A_9$ & $2C_5$ & $1$ & $0$ & $0$ & $1$ & $\frac{\omega_3}{5}$ & $\omega_1-\frac{3\omega_3}{5}$\\ \hline
        $57$ & $A_1 + 2A_4 + A_9$ & $C_1 + A_4 + C_5$ & $1$ & $0$ & $0$ & $1$ & $\frac{\omega_3}{5}$ & $0$\\ \hline
        $81$ & $3A_1 + 2A_2 + A_{11}$ & $A_1 + C_1 + A_2 + C_6$ & $1$ & $0$ & $0$ & $1$ & $\frac{\omega_2}{6}$ & $0$\\ \hline
        $83$ & $2A_2 + A_3 + A_{11}$ & $A_2 + C_2 + C_6$ & $1$ & $0$ & $0$ & $1$ & $\frac{\omega_6}{2}$ & $\frac{\omega_3}{3}-\frac{5\omega_6}{6}$\\ \hline
        $87$ & $2A_1 + A_5 + A_{11}$ & $A_1 + C_3 + C_6$ & $1$ & $0$ & $0$ & $1$ & $\frac{\omega_2}{6}$ & $w_7-\frac{2\omega_2}{3}$\\ \hline
        $106$ & $3A_1 + A_{15}$ & $A_1 + C_1 + C_8$ & $1$ & $0$ & $0$ & $1$ & $\frac{\omega_7}{4}$ & $0$\\ \hline
        $108$ & $A_3 + A_{15} $ & $C_2 + C_8$ & $1$ & $0$ & $0$ & $1$ & $\frac{\omega_6}{2}$ & $\frac{\omega_1}{2}-\frac{3\omega_6}{4}$\\ \hline
        $111$ & $A_1 + A_{17}$ & $C_1 + C_9$ & $1$ & $0$ & $0$ & $1$ & $\frac{\omega_1}{3}$ & $0$\\ \hline
        $121$ & $A_1 + A_7 + 2D_5$ & $C_1 + C_4 +D_5
        $ & $1$ & $0$ & $0$ & $1$ & $\frac{\omega_4}{4}$ & $0$\\ \hline
        $137$ & $2A_3 + 2D_6$ & $2C_2 + D_6$ & $1$ & $0$ & $0$ & $1$ & $\frac{\omega_6}{2}$ & $\frac{\omega_6}{2}-\frac{\omega_8}{2}$\\ \hline
        $219$ & $3E_6$ & $E_6 + F_4$ & $1$ & $-1$ & $0$ & $1$ & $\frac{\omega_5}{3}$ & $0$\\ \hline
        $222$ & $A_1 + A_5 + 2E_6$ & $C_1 + C_3 + E_6$ & $1$ & $0$ & $0$ & $1$ & $\frac{\omega_5}{3}$ & $0$\\ \hline
        $257$ & $A_1 + A_3 + 2E_7$ & $C_1 + C_2 + E_7$ & $1$ & $0$ & $0$ & $1$ & $\frac{\omega_6}{2}$ & $0$\\ \hline
        $279$ & $D_4 + 2E_7$ & $B_3 + E_7$ & $1$ & $-1$ & $0$ & $1$ & $\frac{\omega_6}{2}$ & $0$\\ \hline
        $296$ & $2A_1 + 2E_8$ & $2C_1 + E_8$ & $1$ & $0$ & $0$ & $1$ & $0$ & $0$\\ \hline
        $297$ & $A_2 + 2E_8$ & $A_2+E_8$ & $1$ & $-1$ & $0$ & $1$ & $0$ & $0$\\ \hline
    \end{tabular}
    \caption{Maximal enhancements and corresponding moduli in the $E_8\times E'_8$ and $E_8$ strings on $T^2$, where Wilson lines $a_1,a_2$ are expressed by data in Table \ref{table_E8_simple_roots}.  All of them except for $\#1,2$ can be obtained with the method of generalized Dynkin diagrams.}
    \label{table_maximal_enhancement_T2}
\end{table}

\subsubsection{Massless Matters}
As in the case with 9d, the massless conditions for the co-spinor states are given by \eqref{massless condition2} with $P_{L}^2=1$. Solving these conditions with the corresponding moduli, we can obtain the same results in Table \ref{tab:8d_summary}.

\subsection{Generalized Dynkin Diagram for \texorpdfstring{$T^2$}{T2}}\label{sec:8d_Dynkin}
We propose the procedure for identifying the maximal enhanced gauge groups in the $E_8$ string on $T^2$, which is similar to the method with generalized Dynkin diagrams proposed in the SUSY theories. For the $T^2$ case, the charge vector is generalized as
\begin{align}
    Z=\ket{2w_1,2w_2,n_1,n_2;\pi_{+}},
\end{align}
and the metric is expressed as
\begin{equation}
    \eta=
    \begin{pmatrix}
        0 & \boldsymbol{1}_{2} & 0 \\
        \boldsymbol{1}_{2} & 0 & 0 \\
        0 & 0 & \boldsymbol{1}_{8}
    \end{pmatrix}.
\end{equation}
Note again $Z^2=\pi_{+}^2+4w^1n_1+4w^2n_2=2P^2$. As discussed in \cite{Font:2020rsk}, to obtain the maximal enhancements, we take the Wilson line as $a_1=\omega_k/\kappa_k$, where $k$ is the number of the deleted node. The subgroup left invariant by $a_1$ is denoted by $\mathrm{H}_{k}$, which can be obtained from Figure \ref{Fig_9d_EDD} by deleting the node corresponding to the root $\alpha_{k}$. For $a_2$, which breaks $\mathrm{H}_{k}$ further, we use the shift algorithm and consider the following procedures: we append the one affine root $\hat{\alpha}_{k}$ of the subgroup $\mathrm{H}_{k}$, and then delete one node corresponding to the root $\alpha_{p}$ with $p\neq k$ of the subgroup. The two Wilson lines $a_1,a_2$ finally leave a subgroup of $E_8$ unbroken and the surviving simple roots are $\alpha_{i},~i\neq k,p$ and $\hat{\alpha}_{k}$. By the shift algorithm, we can obtain $a_2$ by the following conditions:
\begin{align}\label{condition_WL}
    a_2\cdot\alpha_i=0,~i\neq k,p,~~~~a_2\cdot\hat{\alpha}_k=-1.
\end{align}
The affine root $\hat{\alpha}_{k}$ can be expressed as
\begin{align}
    \hat{\alpha}_{k}=-\sum_{i\neq k}\hat{\kappa}_{i}\alpha_{i},
\end{align}
where $\hat{\kappa}_{i}$ is the Kac mark of the $i$-th node for the Dynkin diagram of $\mathrm{H}_{k}$. Then from \eqref{condition_WL}, the Wilson line $a_2$ is given by
\begin{align}\label{WL_a2}
    a_2=\frac{\kappa_p}{\hat{\kappa}_p}\left( \frac{\omega_p}{\kappa_p}-\frac{\omega_k}{\kappa_k} \right).
\end{align}

The nodes 0 through 8 are simply generalized to
\begin{align}
    Z_{i}&=\ket{0,0,0,0;\alpha_{i}},~~~i=1,\ldots,8,\\
    Z_{0}&=\ket{0,0,-1,0;\alpha_{0}}.
\end{align}
The new affine node associated with $\hat{\alpha}_{k}$ is
\begin{align}
    Z_{-1}=\ket{0,0,0,-1;\hat{\alpha}_{k}}.
\end{align}
We need to choose the tensor $E_{ij}$ in $T^2$ case to fix the $\texttt{C}$ node. We focus on the two choices as in \cite{Font:2020rsk}:
\begin{align}
    E_1=\begin{pmatrix}
        1 & 0  \\
        0 & 1 
    \end{pmatrix},~~~
    E_2=\begin{pmatrix}
        1 & -1  \\
        0 & 1 
    \end{pmatrix}.
\end{align}
For $E=E_1$, we obtain the two charged vectors
\begin{align}
    Z_{\texttt{C}_1}=\ket{2,0,1,0;0^8},
    ~~~Z_{\texttt{C}_2}=\ket{0,2,0,1;0^8},
\end{align}
and for $E=E_2$,
\begin{align}
    Z_{\texttt{C}_1}=\ket{2,0,1,0;0^8},
    ~~~Z_{\texttt{C}_3}=\ket{0,2,-1,1;0^8}.
\end{align}
Note that these nodes correspond to long roots since $Z^2=2P^2=4$. From their inner products, one can see $Z_{\texttt{C}_1}$ and $Z_{\texttt{C}_2}$ are disconnected while $Z_{\texttt{C}_1}$ and $Z_{\texttt{C}_3}$ are connected.

The procedures to obtain the maximal enhancements using generalized Dynkin diagrams are summarized as follows:
\begin{itemize}
    \item We start from the generalized Dynkin diagram for the $E_8$ string theory on $S^1$ shown in Figure \ref{Fig_9d_EDD} and omit the $k$-th node using the Wilson line $a_1=\omega_{k}/\kappa_{k}$. 
    \item Next we append the affine node $Z_{-1}$ to the ADE types of the disconnected diagram, and furthermore connect the node $Z_{\texttt{C}_2}$ or $Z_{\texttt{C}_3}$ to the affine one.
    \item Finally we delete the $p$-th node using the Wilson line $a_2$ given by \eqref{WL_a2} so that the remaining diagrams are Dynkin diagrams.
\end{itemize}
The diagrams appeared in this procedures can be obtained by folding the ones in the procedures for the SUSY theories.

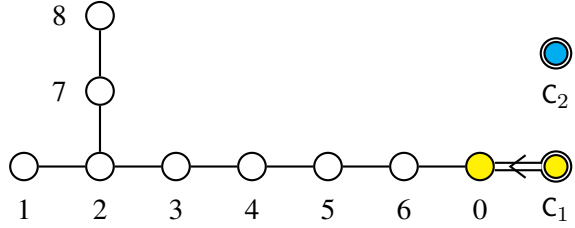
\begin{figure}[htb]
    \begin{center}
        \begin{tikzpicture}[scale=1.0]
            \tikzset{double circle/.style={draw, circle, double, thick, minimum size=10pt, inner sep=1pt}}
            
    \node[thick, circle, draw, fill=white] (1) at (0,0) {};
    \node[thick, circle, draw, fill=white] (2) at (1,0) {};
    \node[thick, circle, draw, fill=white] (3) at (2,0) {};
    \node[thick, circle, draw, fill=white] (4) at (3,0) {};
    \node[thick, circle, draw, fill=white] (5) at (4,0) {};
    \node[thick, circle, draw, fill=white] (6) at (5,0) {};
    \node[thick, circle, draw, fill=white] (7) at (1,1) {};
    \node[thick, circle, draw, fill=white] (8) at (1,2) {};
    \node[thick, circle, draw, fill=yellow] (0) at (6,0) {};
    \node[double circle, fill=yellow] (C1) at (7,0) {};
    \node[double circle, fill=cyan] (C2) at (7,1.5) {};

    \draw[thick] (1) -- (2);
    \draw[thick] (2) -- (3);
    \draw[thick] (3) -- (4);
    \draw[thick] (4) -- (5);
    \draw[thick] (5) -- (6);
    \draw[thick] (2) -- (7);
    \draw[thick] (7) -- (8);
    \draw[thick] (6) -- (0);
    \draw[double distance=2pt, thick] (0) -- (C1);
    \draw[thick] (6.6,0.15) -- (6.4,0.0) -- (6.6,-0.15);

    \node[below=0.3cm] at (1) {1};
    \node[below=0.3cm] at (2) {2};
    \node[below=0.3cm] at (3) {3};
    \node[below=0.3cm] at (4) {4};
    \node[below=0.3cm] at (5) {5};
    \node[below=0.3cm] at (6) {6};
    \node[left=0.3cm] at (7) {7};
    \node[left=0.3cm] at (8) {8};
    \node[below=0.3cm] at (0) {0};
    \node[below=0.3cm] at (C1) {$\texttt{C}_{1}$};
    \node[below=0.3cm] at (C2) {$\texttt{C}_{2}$};
        \end{tikzpicture}
        \caption{Generalized Dynkin diagram for 8d theories with $E=E_1$ and $a_2=0$.}\label{Fig_8d_EDD_E1a0}
    \end{center}
\end{figure}

Let us see some examples. We take the simplest case with $E=E_1, a_2=0$ as a first example. Since $a_2$ vanishes, the node $Z_{-1}$ is deleted. The generalized Dynkin diagram in this case is shown in Figure \ref{Fig_8d_EDD_E1a0}. Using $a_1=\omega_{i}/\kappa_{i},~i=0,\ldots,7$, we obtain eight allowed maximal enhancements by deleting the $i$-th node, which correspond to $\#57,81,106,111,121,222,257,296$. Note that the $8$ node cannot be omitted because it does not lead to Dynkin diagrams.

\begin{figure}[htb]
    \begin{center}
        \begin{tikzpicture}[scale=1.0]
            \tikzset{double circle/.style={draw, circle, double, thick, minimum size=10pt, inner sep=1pt}}
            
    \node[thick, circle, draw, fill=white] (1) at (0,0) {};
    \node[thick, circle, draw, fill=white] (2) at (1,0) {};
    \node[thick, circle, draw, fill=white] (3) at (2,0) {};
    \node[thick, circle, draw, fill=white] (4) at (3,0) {};
    \node[thick, circle, draw, fill=white] (5) at (4,0) {};
    \node[thick, circle, draw, fill=white] (6) at (5,0) {};
    \node[thick, circle, draw, fill=white] (7) at (1,1) {};
    \node[thick, circle, draw, fill=white] (8) at (1,2) {};
    \node[thick, circle, draw, fill=yellow] (0) at (6,0) {};
    \node[double circle, fill=yellow] (C1) at (7,0) {};
    \node[double circle, fill=cyan] (C2) at (7,1) {};

    \draw[thick] (1) -- (2);
    \draw[thick] (2) -- (3);
    \draw[thick] (3) -- (4);
    \draw[thick] (4) -- (5);
    \draw[thick] (5) -- (6);
    \draw[thick] (2) -- (7);
    \draw[thick] (7) -- (8);
    \draw[thick] (6) -- (0);
    \draw[double distance=2pt, thick] (0) -- (C1);
    \draw[thick] (6.6,0.15) -- (6.4,0.0) -- (6.6,-0.15);
    \draw[thick] (C1) -- (C2);

    \node[below=0.3cm] at (1) {1};
    \node[below=0.3cm] at (2) {2};
    \node[below=0.3cm] at (3) {3};
    \node[below=0.3cm] at (4) {4};
    \node[below=0.3cm] at (5) {5};
    \node[below=0.3cm] at (6) {6};
    \node[left=0.3cm] at (7) {7};
    \node[left=0.3cm] at (8) {8};
    \node[below=0.3cm] at (0) {0};
    \node[below=0.3cm] at (C1) {$\texttt{C}_{1}$};
    \node[left=0.3cm] at (C2) {$\texttt{C}_{3}$};
        \end{tikzpicture}
        \caption{Generalized Dynkin diagram for 8d theories with $E=E_2$ and $a_2=0$.}\label{Fig_8d_EDD_E2a0}
    \end{center}
\end{figure}
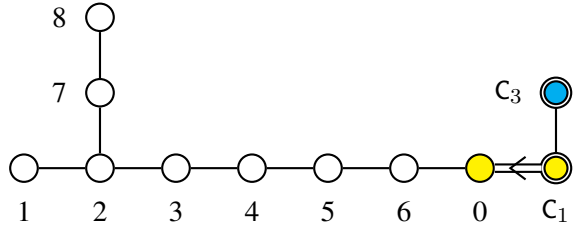

Next we take the case with $E=E_2, a_2=0$. The difference from the first example is that the node $Z_{\texttt{C}_3}$ appears instead of $Z_{\texttt{C}_2}$ and it is connected to the node $Z_{\texttt{C}_1}$. The generalized Dynkin diagram in this case is shown in Figure \ref{Fig_8d_EDD_E2a0}. As in the case with $E=E_1, a_2=0$, we obtain the allowed maximal enhancements by specific $a_1$. However, in this case, The nodes that can be deleted are more restricted. To get possible Dynkin diagrams, we have only three choices: the $0$-, $5$- or $6$-th node that we can omit. These correspond to $\#219,279,297$. Note that the moduli $E=E_2$, $a_1\neq0, a_2\neq0$ cannot lead to any maximal enhancements because deleting any nodes cannot lead to Dynkin diagrams, as in the case of the SUSY theories.

\begin{figure}[htb]
    \begin{center}
        \begin{tikzpicture}[scale=1.0]
            \tikzset{double circle/.style={draw, circle, double, thick, minimum size=10pt, inner sep=1pt}}
            
    \node[thick, circle, draw, fill=white] (1) at (0,0) {};
    \node[thick, circle, draw, fill=white] (2) at (1,0) {};
    \node[thick, circle, draw, fill=white] (3) at (2,0) {};
    \node[thick, circle, draw, fill=white] (4) at (3,0) {};
    \node[thick, circle, draw, fill=white] (5) at (4,0) {};
    \node[thick, circle, draw, fill=white] (7) at (1,1) {};
    \node[thick, circle, draw, fill=white] (8) at (1,2) {};
    \node[thick, circle, draw, fill=yellow] (0) at (6,0) {};
    \node[double circle, fill=yellow] (C1) at (7,0) {};
    \node[double circle, fill=cyan] (C2) at (7,3) {};
    \node[double circle, fill=cyan] (-1) at (1,3) {};

    \draw[thick] (1) -- (2);
    \draw[thick] (2) -- (3);
    \draw[thick] (3) -- (4);
    \draw[thick] (4) -- (5);
    \draw[thick] (2) -- (7);
    \draw[thick] (7) -- (8);
    \draw[double distance=2pt, thick] (0) -- (C1);
    \draw[thick] (6.6,0.15) -- (6.4,0.0) -- (6.6,-0.15);
    \draw[thick] (-1) -- (8);
    \draw[double distance=2pt, thick] (-1) -- (C2);
    \draw[thick] (4.1,3.15) -- (3.9,3.0) -- (4.1,3.0-0.15);

    \node[below=0.3cm] at (1) {1};
    \node[below=0.3cm] at (2) {2};
    \node[below=0.3cm] at (3) {3};
    \node[below=0.3cm] at (4) {4};
    \node[below=0.3cm] at (5) {5};
    \node[left=0.3cm] at (7) {7};
    \node[left=0.3cm] at (8) {8};
    \node[below=0.3cm] at (0) {0};
    \node[below=0.3cm] at (C1) {$\texttt{C}_{1}$};
    \node[below=0.3cm] at (C2) {$\texttt{C}_{2}$};
    \node[left=0.3cm] at (-1) {-1};
        \end{tikzpicture}
        \caption{Generalized Dynkin diagram corresponding to the choice $E=E_1, a_1=\omega_6/2$.}\label{Fig_8d_EDD_E1w6}
    \end{center}
\end{figure}
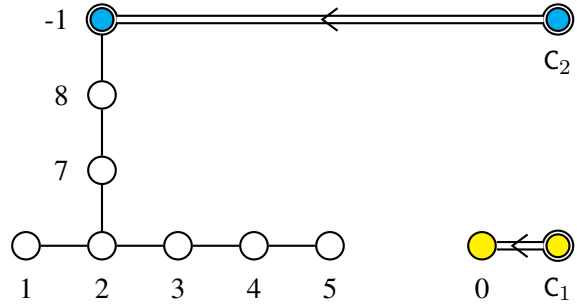

As the last example, we take the case with $E=E_1, a_1=\omega_6/2$. The generalized Dynkin diagram in this case is shown in Figure \ref{Fig_8d_EDD_E1w6}. Using $a_2$ given by \eqref{WL_a2} with $k=6,~p=1,3,6,8$, we obtain four allowed maximal enhancements by deleting the $i$-th node, which correspond to $\#83,108,137,$ and again $\#257$.

Similarly, the other six maximal enhancements $\#5,6,25,28,54,87$ can also be obtained by the above procedures.

\acknowledgments
We thank Justin Kaidi for useful discussions.
The work of Y.H. was supported by MEXT Leading Initiative for Excellent Young Researchers Grant No.JPMXS0320210099, JSPS KAKENHI Grants No.24H00976, 24K07035, and 24KF0167. The work of Y.K. was supported in part by the Inamori Foundation through the Institute for Advanced Study at Kyushu University.

\newpage
\begin{appendix}

\section{Theta functions and \texorpdfstring{$D_4$}{D4} characters}\label{sec:theta}
In this appendix, we summarize the formula used in the paper.
Let $\tau$ be a complex number with positive imaginary part, and $q=\exp \left(2\pi i \tau\right)$. Theta functions are 

\begin{equation}
\begin{aligned}
& \theta_1(\tau)\coloneqq i \sum_{n \in \mathbb{Z}}(-1)^n q^{\frac{1}{2}\left(n-\frac{1}{2}\right)^2}=0, \\
& \theta_2(\tau)\coloneqq\sum_{n \in \mathbb{Z}} q^{\frac{1}{2}\left(n-\frac{1}{2}\right)^2},  \\
& \theta_3(\tau)\coloneqq\sum_{n \in \mathbb{Z}} q^{\frac{1}{2} n^2}, \\
& \theta_4(\tau)\coloneqq\sum_{n \in \mathbb{Z}}(-1)^n q^{\frac{1}{2} n^2}.
\end{aligned}
\end{equation}
They can also be expressed as infinite products:
\begin{equation}
\begin{aligned}
& \theta_2(\tau)=2 q^{\frac{1}{8}} \prod_{m=1}^{\infty}\left(1-q^m\right)\left(1+ q^m\right)\left(1+ q^m\right), \\
& \theta_3(\tau)=\prod_{m=1}^{\infty}\left(1-q^m\right)\left(1+ q^{m-\frac{1}{2}}\right)\left(1+ q^{m-\frac{1}{2}}\right), \\
& \theta_4(\tau)=\prod_{n=0}^{\infty}\left(1-q^m\right)\left(1- q^{m-\frac{1}{2}}\right)\left(1-q^{m-\frac{1}{2}}\right) .
\end{aligned}
\end{equation}
Modular transformation properties of theta functions are given as follows:
\begin{equation}
\begin{aligned}
      \qty(\theta_2 ,\theta_3, \theta_4)(\tau+1)=&\qty(e^{\frac{\pi i}{4}}\theta_2,~\theta_4,~\theta_3)(\tau),\\
\begin{pmatrix}
    \theta_2 \\\theta_3 \\\theta_4 
\end{pmatrix}\qty(-\frac{1}{\tau})=&
e^{-\frac{i\pi}{4}}\tau^{\frac{1}{2}}\left(\begin{array}{cccc}
0 & 0 & 1  \\
0 & 1 & 0  \\
1 & 0 & 0 
\end{array}\right)
\begin{pmatrix}
    \theta_2 \\\theta_3 \\\theta_4
\end{pmatrix}(\tau).
\end{aligned}
\label{eq:theta}
\end{equation}
The conjugacy classes are
\begin{equation}
    \Gamma_g^{(4)}=\left\{\left(n_1, \cdots, n_4\right) \mid n_i \in \mathbb{Z}, \sum_{i=1}^4 n_i \in 2 \mathbb{Z}\right\} ,
\end{equation}
\begin{equation}
    \Gamma_v^{(4)}=\left\{\left(n_1, \cdots, n_4\right) \mid n_i \in \mathbb{Z}, \sum_{i=1}^4 n_i \in 2 \mathbb{Z}+1\right\} ,
\end{equation}
\begin{equation}
    \Gamma_s^{(4)}=\left\{\left.\left(n_1+\frac{1}{2}, \cdots, n_4+\frac{1}{2}\right) \right\rvert\, n_i \in \mathbb{Z}, \sum_{i=1}^4 n_i \in 2 \mathbb{Z}\right\} ,
\end{equation}
\begin{equation}
    \Gamma_c^{(4)}=\left\{\left.\left(n_1+\frac{1}{2}, \cdots, n_4+\frac{1}{2}\right) \right\rvert\, n_i \in \mathbb{Z}, \sum_{i=1}^4 n_i \in 2 \mathbb{Z}+1\right\}.
\end{equation}
These lattices give the $D_4$ characters:
\begin{equation}
\begin{aligned}
& O_{8}=\frac{1}{\eta^4} \sum_{p\in \Gamma_g^{(4)}} q^{\frac{1}{2}|\pi|^2}=\frac{1}{2 \eta^4}\left(\theta_3^4(\tau)+\theta_4^4(\tau)\right), \\
& V_{8}=\frac{1}{\eta^4} \sum_{p\in \Gamma_v^{(4)}} q^{\frac{1}{2}|\pi|^2}=\frac{1}{2 \eta^4}\left(\theta_3^4(\tau)-\theta_4^4(\tau)\right), \\
& S_{8}=\frac{1}{\eta^4} \sum_{p\in \Gamma_s^{(4)}} q^{\frac{1}{2}|\pi|^2}=\frac{1}{2 \eta^4}\left(\theta_2^4(\tau)+\theta_1^4(\tau)\right), \\
& C_{8}=\frac{1}{\eta^4} \sum_{p\in \Gamma_c^{(4)}} q^{\frac{1}{2}|\pi|^2}=\frac{1}{2 \eta^4}\left(\theta_2^4(\tau)-\theta_1^4(\tau)\right).
\label{eq:D4_characters}\end{aligned}
\end{equation}
It follows from equation (\ref{eq:theta}) that they transform as 
\begin{equation}
\begin{aligned}
      \qty(O_8 ,V_8,S_8,C_8)(\tau+1)=&\qty(e^{-\frac{1}{3}\pi i}O_8,~-e^{-\frac{1}{3}\pi i}V_8,~e^{\frac{2}{3}\pi i}S_8,~e^{\frac{2}{3}\pi i}C_8)(\tau),\\
\begin{pmatrix}
    O_8 \\V_8 \\S_8 \\C_8
\end{pmatrix}\qty(-\frac{1}{\tau})=&\frac{1}{2}\left(\begin{array}{cccc}
1 & 1 & 1 & 1 \\
1 & 1 & -1 & -1 \\
1 & -1 & 1 & -1 \\
1 & -1 & -1 & 1
\end{array}\right)\begin{pmatrix}
    O_8 \\V_8 \\S_8 \\C_8
\end{pmatrix}(\tau).
\end{aligned}
\label{eq:fermions_modular_tr}\end{equation}
The small $q$ expansions are
\begin{equation}
    \begin{aligned}
        &O_8=\frac{1}{\eta^4}\qty(1+24q+\cdots),
        &&V_8=\frac{1}{\eta^4}\qty(8q^{\frac{1}{2}}+\cdots),\\
        &S_8=\frac{1}{\eta^4}\qty(8q^{\frac{1}{2}}+\cdots),
        &&C_8=\frac{1}{\eta^4}\qty(8q^{\frac{1}{2}}+\cdots).\\
    \end{aligned}
\end{equation}
The definition of the Dedekind eta function is
\begin{equation}
    \eta(\tau)=q^{\frac{1}{24}}\prod_{n=1}^\infty \qty(1-q^n).
\label{eq:eta_function}\end{equation}
The modular transformation properties are
\begin{equation}
    \begin{aligned}
        &\eta(\tau+1)=e^{\frac{\pi i}{12}}\eta(\tau),
        &&\eta\qty(-\frac{1}{\tau})=\sqrt{-i\tau}\eta(\tau).
    \end{aligned}
\label{eq:eta_modular_tr}\end{equation}

    \section{Root and Weight Lattices}\label{sec:lie algebra}
In this appendix, we summarize the details of the root lattices and weight lattices used in section \ref{sec:orbifold}, \ref{sec:8d_folding} and \ref{sec:E8}. For notation and other details see \cite{Bourbaki:2002}. Note that for $E_8$, we use the same natation as in \cite{Font:2020rsk}.
We denote the $i$-th standard orthonormal basis of $\R^n$ by $\varepsilon_i$.

\subsection{\texorpdfstring{$A_n$}{An} type}

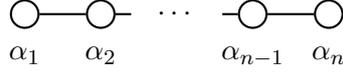
\begin{figure}[htb]
    \begin{center}
        \begin{tikzpicture}[scale=1.0]
            \node[thick, circle, draw, fill=white] (1) at (0,0) {};
            \node[thick, circle, draw, fill=white] (2) at (1,0) {};
            \node at (2,0) {$\cdots$};
            \node[thick, circle, draw, fill=white] (n-1) at (3,0) {};
            \node[thick, circle, draw, fill=white] (n) at (4,0) {};

            \draw[thick] (1) -- (2);
            \draw[thick] (2) -- (1.4,0);
            \draw[thick] (2.6,0) -- (n-1);
            \draw[thick] (n-1) -- (n);

            \node[below=0.3cm] at (1) {$\alpha_1$};
            \node[below=0.3cm] at (2) {$\alpha_2$};
            \node[below=0.3cm] at (n-1) {$\alpha_{n-1}$};
            \node[below=0.3cm] at (n) {$\alpha_n$};
        \end{tikzpicture}
        \caption{Dynkin diagram of type $A_n$.}
        \label{fig:An_Dynkin}
    \end{center}
\end{figure}

The Dynkin diagram of the $A_n$ algebra is shown in figure~\ref{fig:An_Dynkin}.
The root system of $A_n$ is
        \begin{equation}
    \varepsilon_j-\varepsilon_k, 1\leq j<k\leq n+1, ~j\neq k.
\end{equation}
The basis are
        \begin{equation}
\begin{aligned}
        &\alpha_i=\varepsilon_i-\varepsilon_{i+1},&&\text{for}\quad 1\leq i \leq n.
\end{aligned}
\end{equation}
The fundamental weights of $A_n$ are
    \begin{equation}
        \begin{aligned}
            \omega_i=&\varepsilon_1+\cdots+\varepsilon_i-\frac{i}{n+1}(\varepsilon_1+\cdots+\varepsilon_{n+1})\\
            =&\frac{1}{n+1}\Bigl[(n-i+1)(\alpha_1+2\alpha_2+\cdots + (i-1)\alpha_{i-1})\\
            +&i\bigl((n-i+1)\alpha_i+(n-i)\alpha_{i+1}+\cdots+\alpha_n\bigr)\Bigr].
        \end{aligned}
    \end{equation}


\subsection{\texorpdfstring{$B_n$}{Bn} type}
\begin{figure}[htb]
    \begin{center}
        \begin{tikzpicture}[scale=1.0]
            \node[thick, circle, draw, fill=white] (1) at (0,0) {};
            \node[thick, circle, draw, fill=white] (2) at (1,0) {};
            \node at (2,0) {$\cdots$};
            \node[thick, circle, draw, fill=white] (n-1) at (3,0) {};
            \node[thick, circle, draw, fill=white] (n) at (4,0) {};

            \draw[thick] (1) -- (2);
            \draw[thick] (2) -- (1.4,0);
            \draw[thick] (2.6,0) -- (n-1);
            \draw[double distance=2pt, thick] (n-1) -- (n);
            \draw[thick] (3.45,0.15) -- (3.55,0) -- (3.45,-0.15); 

            \node[below=0.3cm] at (1) {$\alpha_1$};
            \node[below=0.3cm] at (2) {$\alpha_2$};
            \node[below=0.3cm] at (n-1) {$\alpha_{n-1}$};
            \node[below=0.3cm] at (n) {$\alpha_n$};
        \end{tikzpicture}
        \caption{Dynkin diagram of type $B_n$.}
        \label{fig:Bn_Dynkin}
    \end{center}
\end{figure}
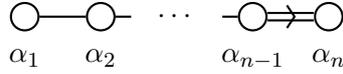

The Dynkin diagram of the $B_n$ algebra is shown in figure~\ref{fig:Bn_Dynkin}.
The root system of $B_n$ is 
\begin{equation}
    \begin{aligned}
        \pm\varepsilon_i,1\leq i\leq n,\\ \pm\varepsilon_i\pm\varepsilon_j,1\leq i<j\leq n
    \end{aligned}
\end{equation}
The fundamental weights of $B_n$ are
\begin{equation}
    \begin{aligned}
        \omega_{i\neq n}=&\varepsilon_1+\cdots+\varepsilon_i\\
        =&\alpha_1+2\alpha_2+\cdots+(i-1)\alpha_{i-1}+i(\alpha_i+\cdots +\alpha_n),\\
        \omega_n=&\frac{1}{2}(\varepsilon_1+\cdots+\varepsilon_n)\\
        =&\frac{1}{2}\qty(\alpha_1+2\alpha_2+\cdots+ n\alpha_n).
        \end{aligned}
\end{equation}

\subsection{\texorpdfstring{$C_n$}{Cn} type}
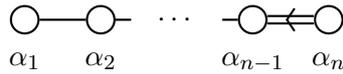
\begin{figure}[htb]
    \begin{center}
        \begin{tikzpicture}[scale=1.0]
            \node[thick, circle, draw, fill=white] (1) at (0,0) {};
            \node[thick, circle, draw, fill=white] (2) at (1,0) {};
            \node at (2,0) {$\cdots$};
            \node[thick, circle, draw, fill=white] (n-1) at (3,0) {};
            \node[thick, circle, draw, fill=white] (n) at (4,0) {};

            \draw[thick] (1) -- (2);
            \draw[thick] (2) -- (1.4,0);
            \draw[thick] (2.6,0) -- (n-1);
            \draw[double distance=2pt, thick] (n-1) -- (n);
            \draw[thick] (3.55,0.15) -- (3.45,0) -- (3.55,-0.15); 

            \node[below=0.3cm] at (1) {$\alpha_1$};
            \node[below=0.3cm] at (2) {$\alpha_2$};
            \node[below=0.3cm] at (n-1) {$\alpha_{n-1}$};
            \node[below=0.3cm] at (n) {$\alpha_n$};
        \end{tikzpicture}
        \caption{Dynkin diagram of type $C_n$.}
        \label{fig:Cn_Dynkin}
    \end{center}
\end{figure}
The Dynkin diagram of the $C_n$ algebra is shown in figure~\ref{fig:Cn_Dynkin}.
The root system of $C_n$ is
\begin{equation}
\begin{aligned}
         &\pm2\varepsilon_j, 1\leq j \leq n,\\
     &\pm \varepsilon_j\pm \varepsilon_k , 1\leq j<k\leq n.
\end{aligned}
\end{equation}
The basis of the $C_n$ root is
\begin{equation}
\begin{aligned}
        &\alpha_i=\varepsilon_i-\varepsilon_{i+1},&&\text{for}\quad1\leq i \leq n-1 ,\\
        &\alpha_n=2\varepsilon_n.
\end{aligned}
\end{equation}
The fundamental weights of $C_n$ are
    \begin{equation}
\begin{aligned}
\omega_i= & \varepsilon_1+\varepsilon_2+\cdots+\varepsilon_i \\
=&\alpha_1+2\alpha_2 +\cdots+(i-1) \alpha_{i-1} \\
+&i\left(\alpha_i+\alpha_{i+1}+\cdots+\alpha_{n-1}+\frac{1}{2} \alpha_n\right),
\end{aligned}
\end{equation}
The root lattice and weight lattice have the following relationships:
\begin{equation}
    \begin{aligned}
        (\Lambda_R(C_n))^\ast=&\Lambda_W(C_n)+\Z\frac{1}{2}\omega_n\\
        =&\Lambda_R(C_n)+\Z\frac{1}{2}\alpha_n+\Z\frac{1}{2}\omega_n^{(C_n)},\\
        & (\Lambda_R(C_n)+\frac{1}{2}\Z\alpha_n^{(C_n)})^\ast=(\Z^n)^\ast\\
        =&\Lambda_R(C_n)+\frac{1}{2}\Z\alpha_n^{(C_n)}\\
    \Lambda_W(C_n)/&\Lambda_R(C_n)=\Z_{2}.
    \end{aligned}
\end{equation}

\subsection{\texorpdfstring{$D_n$}{Dn} type}
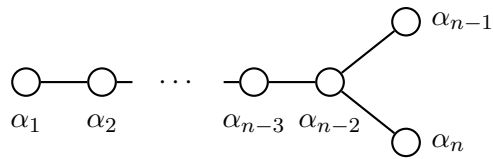
\begin{figure}[htb]
    \begin{center}
        \begin{tikzpicture}[scale=1.0]
            \node[thick, circle, draw, fill=white] (1) at (0,0) {};
            \node[thick, circle, draw, fill=white] (2) at (1,0) {};
            \node at (2,0) {$\cdots$};
            \node[thick, circle, draw, fill=white] (n-3) at (3,0) {};
            \node[thick, circle, draw, fill=white] (n-2) at (4,0) {};

            \node[thick, circle, draw, fill=white] (n-1) at (5,0.8) {};
            \node[thick, circle, draw, fill=white] (n)   at (5,-0.8) {};

            \draw[thick] (1) -- (2);
            \draw[thick] (2) -- (1.4,0);
            \draw[thick] (2.6,0) -- (n-3);
            \draw[thick] (n-3) -- (n-2);
            \draw[thick] (n-2) -- (n-1);
            \draw[thick] (n-2) -- (n);

            \node[below=0.3cm] at (1) {$\alpha_1$};
            \node[below=0.3cm] at (2) {$\alpha_2$};
            \node[below=0.3cm] at (n-3) {$\alpha_{n-3}$};
            \node[below=0.3cm] at (n-2) {$\alpha_{n-2}$};
            \node[right=0.2cm] at (n-1) {$\alpha_{n-1}$};
            \node[right=0.2cm] at (n) {$\alpha_n$};
        \end{tikzpicture}
        \caption{Dynkin diagram of type $D_n$.}
        \label{fig:Dn_Dynkin}
    \end{center}
\end{figure}

The Dynkin diagram of the $D_n$ algebra is shown in figure~\ref{fig:Dn_Dynkin}.
The root system of $D_n$ is
\begin{equation}  \pm\varepsilon_j\pm\varepsilon_k,\pm\varepsilon_j\mp\varepsilon_k,\quad 1\leq j<k\leq n.
\end{equation}
The basis are
\begin{equation}
\begin{aligned}
        &\alpha_i=\varepsilon_i-\varepsilon_{i+1},&&\text{for}\quad 1\leq i\leq n-1, \\
        &\alpha_n=\varepsilon_{n-1}+\varepsilon_n.
\end{aligned}
\end{equation}
The fundamental weights of $D_n$ are   
\begin{equation}
\begin{aligned}
\omega_i= & \varepsilon_1+\varepsilon_2+\cdots+\varepsilon_i\\
= & \alpha_1+2 \alpha_2+\cdots+(i-1) \alpha_{i-1}+i\left(\alpha_n+\alpha_{i+1}+\cdots+\alpha_{n-2}\right) \\
& +\frac{1}{2} i\left(\alpha_{n-1}+\alpha_n\right) , \quad\quad \text{for}\quad1 \leq i \leq n-2, \\
\omega_{n-1}= & \frac{1}{2}(\varepsilon_n+\varepsilon_2+\cdots+\varepsilon_{n-2}+\varepsilon_{n-1}-\varepsilon_n) \\
= & \frac{1}{2}\left(\alpha_1+2 \alpha_2+\cdots+(n-2) \alpha_{n-2}+\frac{1}{2} n \alpha_{n-1}+\frac{1}{2}(n-2) \alpha_n\right) ,\\
\omega_n= & \frac{1}{2}\left(\varepsilon_1+\varepsilon_2+\cdots+\varepsilon_{n-2}+\varepsilon_{n-1}+\varepsilon_n\right) \\
= & \frac{1}{2}\qty(\alpha_1+2 \alpha_2+\cdots+(n-2) \alpha_{n-2}+\frac{1}{2}(n-2) \alpha_{n-1}+\frac{1}{2} n\alpha_n).
\end{aligned}
\end{equation}
The root lattice and weight lattice have following relationships:

\begin{equation}
    \begin{aligned}
    \qty(\Lambda_R(D_{2n}))^\ast=&\Lambda_W(D_{2n})\\
    =&\Lambda_R(D_{2n})+\Z\omega_{2n-1}+\Z\omega_{2n},\\
    \qty(\Lambda_R(D_{2n+1}))^\ast=&\Lambda_W(D_{2n+1})\\
    =&\Lambda_R(D_{2n})+\Z\omega_{2n+1},\\
    \Lambda_{W}(D_{2n})/\Lambda_{R}(D_{2n})=\Z_2&\times\Z_2,
    \quad\quad\Lambda_{W}(D_{2n+1})/\Lambda_{R}(D_{2n+1})=\Z_4.
    \end{aligned}
\end{equation}

\subsection{\texorpdfstring{$E_6$}{E6}}
\begin{figure}[htb]
    \begin{center}
        \begin{tikzpicture}[scale=1.0]
            \node[thick, circle, draw, fill=white] (1) at (0,0) {};
            \node[thick, circle, draw, fill=white] (3) at (1,0) {};
            \node[thick, circle, draw, fill=white] (4) at (2,0) {};
            \node[thick, circle, draw, fill=white] (5) at (3,0) {};
            \node[thick, circle, draw, fill=white] (6) at (4,0) {};
            \node[thick, circle, draw, fill=white] (2) at (2,-1) {};

            \draw[thick] (1) -- (3);
            \draw[thick] (3) -- (4);
            \draw[thick] (4) -- (5);
            \draw[thick] (5) -- (6);
            \draw[thick] (4) -- (2); 

            \node[above=0.3cm] at (1) {$\alpha_1$};
            \node[above=0.3cm] at (3) {$\alpha_3$};
            \node[above=0.3cm] at (4) {$\alpha_4$};
            \node[above=0.3cm] at (5) {$\alpha_5$};
            \node[above=0.3cm] at (6) {$\alpha_6$};
 
            \node[right=0.2cm] at (2) {$\alpha_2$};
        \end{tikzpicture}
        \caption{Dynkin diagram of type $E_6$.}
        \label{fig:E6_Dynkin}
    \end{center}
\end{figure}
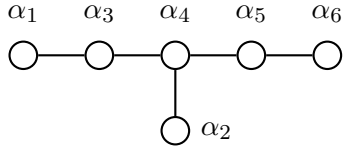
The Dynkin diagram of the $E_6$ algebra is shown in figure~\ref{fig:E6_Dynkin}.
The root system of $E_6$ is
\begin{equation}
\begin{aligned}
    &\pm\varepsilon_i\pm\varepsilon_j, 
    &&\text{for}\quad1\leq i,j\leq 5,\\
    &\pm\frac{1}{2}\qty(\varepsilon_8-\varepsilon_7-\varepsilon_6+\sum_{i=1}^5 (-1)^{\nu_i}\varepsilon_i),
    &&\text{for}\quad\sum_{i=1}^5 \nu_i\in 2\Z.
\end{aligned}
\end{equation}
The basis of $E_6$ root system is
\begin{equation}
\begin{aligned}
&\alpha_1=\frac{1}{2}\left(\varepsilon_1+\varepsilon_8\right)-\frac{1}{2}\left(\varepsilon_2+\varepsilon_3+\varepsilon_4+\varepsilon_5+\varepsilon_6+\varepsilon_7\right), &&\alpha_2=\varepsilon_1+\varepsilon_2, \\
& \alpha_3=\varepsilon_2-\varepsilon_1, &&\alpha_4=\varepsilon_3-\varepsilon_2, \\
&\alpha_5=\varepsilon_4-\varepsilon_3, &&\alpha_6=\varepsilon_5-\varepsilon_4.
\end{aligned}
\end{equation}
The fundamental weights of $E_6$ are
\begin{equation}
\begin{aligned}
\omega_1 =&\frac{2}{3}\left(\varepsilon_8-\varepsilon_7-\varepsilon_6\right)\\
=&\frac{1}{3}\left(4 \alpha_1+3 \alpha_2+5 \alpha_3+6 \alpha_4+4 \alpha_5+2 \alpha_6\right), \\
\omega_2 =&\frac{1}{2}\left(\varepsilon_1+\varepsilon_2+\varepsilon_3+\varepsilon_4+\varepsilon_5-\varepsilon_6-\varepsilon_7+\varepsilon_8\right) \\
=&\alpha_1+2 \alpha_2+2 \alpha_3+3 \alpha_4+2 \alpha_5+\alpha_6,\\
\omega_3  =&\frac{5}{6}\left(\varepsilon_8-\varepsilon_7-\varepsilon_6\right)+\frac{1}{2}\left(-\varepsilon_1+\varepsilon_2+\varepsilon_3+\varepsilon_4+\varepsilon_5\right) \\
=&\frac{1}{3}\left(5 \alpha_1+6 \alpha_2+10 \alpha_3+12 \alpha_4+8 \alpha_5+4 \alpha_6\right) ,\\
\omega_4 =&\varepsilon_3+\varepsilon_4+\varepsilon_5-\varepsilon_6-\varepsilon_7+\varepsilon_8 \\
=&2 \alpha_1+3 \alpha_2+4 \alpha_3+6 \alpha_4+4 \alpha_5+2 \alpha_6, \\
\omega_5 =&\frac{2}{3}\left(\varepsilon_8-\varepsilon_7-\varepsilon_6\right)+\varepsilon_4+\varepsilon_5 \\
=&\frac{1}{3}\left(4 \alpha_1+6 \alpha_2+8 \alpha_3+12 \alpha_4+10 \alpha_5+5 \alpha_6\right), \\
\omega_6 =&\frac{1}{3}\left(\varepsilon_8-\varepsilon_7-\varepsilon_6\right)+\varepsilon_5 \\ 
=&\frac{1}{3}\left(2 \alpha_1+3 \alpha_2+4 \alpha_3+6 \alpha_4+5 \alpha_5+4 \alpha_6\right) .
\end{aligned}
\end{equation}
The relation between the root and weight lattice is
\begin{equation}
    \begin{aligned}
    \qty(\Lambda_R(E_6))^\ast=&\Lambda_W(E_6)\\
    =&\Lambda_R(E_6)+\Z\omega_1,\\
    \Lambda_W(E_6)/\Lambda_R(E_6)=&\Z_3.
    \end{aligned}
\end{equation}

\subsection{\texorpdfstring{$E_7$}{E7}}

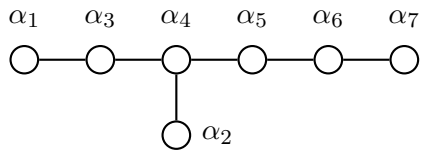
\begin{figure}[htb]
    \begin{center}
        \begin{tikzpicture}[scale=1.0]
            \node[thick, circle, draw, fill=white] (1) at (0,0) {};
            \node[thick, circle, draw, fill=white] (3) at (1,0) {};
            \node[thick, circle, draw, fill=white] (4) at (2,0) {};
            \node[thick, circle, draw, fill=white] (5) at (3,0) {};
            \node[thick, circle, draw, fill=white] (6) at (4,0) {};
            \node[thick, circle, draw, fill=white] (7) at (5,0) {};
            \node[thick, circle, draw, fill=white] (2) at (2,-1) {};

            \draw[thick] (1) -- (3);
            \draw[thick] (3) -- (4);
            \draw[thick] (4) -- (5);
            \draw[thick] (5) -- (6);
            \draw[thick] (6) -- (7);
            \draw[thick] (4) -- (2); 

            \node[above=0.3cm] at (1) {$\alpha_1$};
            \node[above=0.3cm] at (3) {$\alpha_3$};
            \node[above=0.3cm] at (4) {$\alpha_4$};
            \node[above=0.3cm] at (5) {$\alpha_5$};
            \node[above=0.3cm] at (6) {$\alpha_6$};
            \node[above=0.3cm] at (7) {$\alpha_7$};
            \node[right=0.2cm] at (2) {$\alpha_2$};
        \end{tikzpicture}
        \caption{Dynkin diagram of type $E_7$.}
        \label{fig:E7_Dynkin}
    \end{center}
\end{figure}
The Dynkin diagram of the $E_7$ algebra is shown in figure~\ref{fig:E7_Dynkin}.
The root system of $E_7$ is
\begin{equation}
\begin{aligned}
    &\pm\varepsilon_i\pm\varepsilon_j, 
    &&\text{for}\quad1\leq i,j\leq 6,
    \\
    &\pm(\varepsilon_7-\varepsilon_8),\\
    &\pm\frac{1}{2}\qty(\varepsilon_8-\varepsilon_7-\varepsilon_6+\sum_{i=1}^6 (-1)^{\nu_i}\varepsilon_i),
    &&\text{for}\quad\sum_{i=1}^6 \nu_i\in 2\Z.
\end{aligned}
\end{equation}
The basis are
\begin{equation}
\begin{aligned}
& \alpha_1=\frac{1}{2}\left(\varepsilon_1+\varepsilon_8\right)-\frac{1}{2}\left(\varepsilon_2+\varepsilon_3+\varepsilon_4+\varepsilon_5+\varepsilon_6+\varepsilon_7\right), \\
& \alpha_2=\varepsilon_1+\varepsilon_2, \quad\quad\quad\quad\alpha_3=\varepsilon_2-\varepsilon_1, \\
& \alpha_4=\varepsilon_3-\varepsilon_2, \quad\quad\quad\quad \alpha_5=\varepsilon_4-\varepsilon_3, \\ 
& \alpha_6=\varepsilon_5-\varepsilon_4, \quad\quad\quad\quad \alpha_7=\varepsilon_6-\varepsilon_5 .
\end{aligned}
\end{equation}
The fundamental weights of $E_7$ are
\begin{equation}
\begin{aligned}
\omega_1 =&\varepsilon_8-\varepsilon_7\\
=&2 \alpha_1+2 \alpha_2+3 \alpha_3+4 \alpha_4+3 \alpha_5+2 \alpha_6+\alpha_7,\\
\omega_2 =&\frac{1}{2}\left(\varepsilon_1+\varepsilon_2+\varepsilon_3+\varepsilon_4+\varepsilon_5+\varepsilon_6-2 \varepsilon_7+2 \varepsilon_8\right) \\
=&\frac{1}{2}\left(4 \alpha_1+7 \alpha_2+8 \alpha_3+12 \alpha_4+9 \alpha_5+8 \alpha_6+3 \alpha_7\right), \\
\omega_3 =&\frac{1}{2}\left(-\varepsilon_1+\varepsilon_2+\varepsilon_3+\varepsilon_4+\varepsilon_5+\varepsilon_6-3 \varepsilon_7+3 \varepsilon_8\right) \\
=&3 \alpha_1+4 \alpha_2+6 \alpha_3+8 \alpha_4+6 \alpha_5+4 \alpha_6+2 \alpha_7, \\
\omega_4 =&\varepsilon_3+\varepsilon_4+\varepsilon_5+\varepsilon_6+2\left(\varepsilon_8-\varepsilon_7\right) \\
=&4 \alpha_1+6 \alpha_2+8 \alpha_3+12 \alpha_4+9 \alpha_5+6 \alpha_6+3 \alpha_7, \\
\omega_5 =&\varepsilon_4+\varepsilon_5+\varepsilon_6+\frac{3}{2}\left(\varepsilon_8-\varepsilon_7\right) \\
=&\frac{1}{2}\left(6 \alpha_1+9 \alpha_2+12 \alpha_3+18 \alpha_4+15 \alpha_5+10 \alpha_6+5 \alpha_7\right),\\
\omega_6 =&\varepsilon_5+\varepsilon_6-\varepsilon_7+\varepsilon_8 \\
=&2 \alpha_1+3 \alpha_2+4 \alpha_3+6 \alpha_4+5 \alpha_5+4 \alpha_6+2 \alpha_7, \\
\omega_7 =&\varepsilon_6+\frac{1}{2}\left(\varepsilon_8-\varepsilon_7\right) \\
=&\frac{1}{2}\left(2 \alpha_1+3 \alpha_2+4 \alpha_3+6 \alpha_4+5 \alpha_5+4 \alpha_6+3 \alpha_7\right) .
\end{aligned}
\end{equation}
The relation between the root and weight lattice is
\begin{equation}
    \begin{aligned}
    \qty(\Lambda_R(E_7))^\ast=&\Lambda_W(E_7)\\
    =&\Lambda_R(E_7)+\Z\omega_7,\\
    \Lambda_W(E_7)/\Lambda_R(E_7)=&\Z_2.
    \end{aligned}
\end{equation}

\subsection{\texorpdfstring{$E_8$}{E8}}

The basis are

\begin{equation}
    \begin{aligned}
        \alpha_1=&\varepsilon_1-\varepsilon_2,\quad \quad \alpha_2=\varepsilon_2-\varepsilon_3,\\
        \alpha_3=&\varepsilon_3-\varepsilon_4,\quad \quad \alpha_4=\varepsilon_4-\varepsilon_5,\\
        \alpha_5=&\varepsilon_5-\varepsilon_6,\quad \quad \alpha_6=\varepsilon_6-\varepsilon_7,\\
        \alpha_7=&-(\varepsilon_1+\varepsilon_2),\quad\alpha_8=\frac{1}{2}\qty(\varepsilon_1+\cdots+\varepsilon_8).
    \end{aligned}
\end{equation}

\begin{figure}[htb]
    \begin{center}
        \begin{tikzpicture}[scale=1.0]
            \node[thick, circle, draw, fill=white] (1) at (0,0) {};
            \node[thick, circle, draw, fill=white] (3) at (1,0) {};
            \node[thick, circle, draw, fill=white] (4) at (2,0) {};
            \node[thick, circle, draw, fill=white] (5) at (3,0) {};
            \node[thick, circle, draw, fill=white] (6) at (4,0) {};
            \node[thick, circle, draw, fill=white] (7) at (5,0) {};
            \node[thick, circle, draw, fill=white] (8) at (6,0) {};
            \node[thick, circle, draw, fill=white] (2) at (2,-1) {};

            \draw[thick] (1) -- (3);
            \draw[thick] (3) -- (4);
            \draw[thick] (4) -- (5);
            \draw[thick] (5) -- (6);
            \draw[thick] (6) -- (7);
            \draw[thick] (7) -- (8);
            \draw[thick] (4) -- (2); 

            \node[above=0.3cm] at (1) {$\alpha_8$};
            \node[above=0.3cm] at (3) {$\alpha_7$};
            \node[above=0.3cm] at (4) {$\alpha_2$};
            \node[above=0.3cm] at (5) {$\alpha_3$};
            \node[above=0.3cm] at (6) {$\alpha_4$};
            \node[above=0.3cm] at (7) {$\alpha_5$};
            \node[above=0.3cm] at (8) {$\alpha_6$};
            \node[right=0.2cm] at (2) {$\alpha_1$};
        \end{tikzpicture}
        \caption{Dynkin diagram of type $E_8$.}
        \label{fig:E8_Dynkin}
    \end{center}
\end{figure}
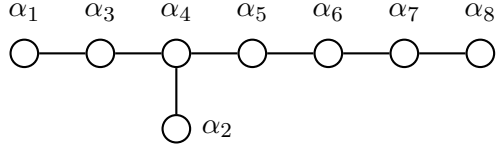

\subsection{\texorpdfstring{$F_4$}{F4}}
\begin{figure}[htb]
    \begin{center}
        \begin{tikzpicture}[scale=1.0]
            \node[thick, circle, draw, fill=white] (1) at (0,0) {};
            \node[thick, circle, draw, fill=white] (2) at (1,0) {};
            \node[thick, circle, draw, fill=white] (3) at (2,0) {};
            \node[thick, circle, draw, fill=white] (4) at (3,0) {};
            
            \draw[thick] (1) -- (2);
            \draw[double distance=2pt, thick] (2) -- (3); 
            \draw[thick] (3) -- (4);

            \draw[thick] (1.45,0.15) -- (1.5,0) -- (1.45,-0.15);

            \node[below=0.3cm] at (1) {$\alpha_1$};
            \node[below=0.3cm] at (2) {$\alpha_2$};
            \node[below=0.3cm] at (3) {$\alpha_3$};
            \node[below=0.3cm] at (4) {$\alpha_4$};
        \end{tikzpicture}
        \caption{Dynkin diagram of type $F_4$.}
        \label{fig:F4_Dynkin}
    \end{center}
\end{figure}
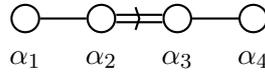

The Dynkin diagram of the $F_4$ algebra is shown in figure~\ref{fig:F4_Dynkin}.
The root system of $F_4$ is
\begin{equation}
\begin{aligned}
        \pm \varepsilon_i, 1\leq i \leq 4,\\
        \pm \varepsilon_i\pm \varepsilon_j, 1\leq i,j \leq 4,\\
        \frac{1}{2}\qty(\pm \varepsilon_1\pm \varepsilon_2\pm \varepsilon_3\pm \varepsilon_4).
\end{aligned}
\end{equation}
The basis of the root system of $F_4$ is
\begin{equation}
\begin{aligned}
      &\alpha_1=\varepsilon_2-\varepsilon_3,
      &&\alpha_2=\varepsilon_3-\varepsilon_4,\\
      &\alpha_3=\varepsilon_4,
      &&\alpha_4=\frac{1}{2}\qty(\varepsilon_1-\varepsilon_2-\varepsilon_3-\varepsilon_4).
\end{aligned}
\end{equation}

\end{appendix}
\bibliographystyle{JHEP}
\bibliography{reference}

\end{document}